\documentclass[useAMS,usenatbib]{mn2e}

\usepackage{amsmath,psfig,epsfig,graphics}
\usepackage{epstopdf}

\def\aap{A\&A}

\def\apj{ApJ}

\def\apjl{ApJ}
\def\mnras{MNRAS}
\def\araa{ARA\&A}

\def\physrep{Phys. Rep.}

\def\aaps{A\&A Supp.}

\def\na{New Astronomy}

\def\pasj{PASJ}

\def\lesssim{\mathrel{\hbox{\rlap{\hbox{\lower4pt\hbox{$\sim$}}}\hbox{$<$}}}}
\def\gesssim{\mathrel{\hbox{\rlap{\hbox{\lower4pt\hbox{$\sim$}}}\hbox{$>$}}}}

\def\lesssim{\mathrel{\hbox{\rlap{\hbox{\lower4pt\hbox{$\sim$}}}\hbox{$<$}}}}
\def\gesssim{\mathrel{\hbox{\rlap{\hbox{\lower4pt\hbox{$\sim$}}}\hbox{$>$}}}}

\begin{document}

\author[Morandi, Nagai, \& Cui]
{Andrea Morandi${}^1$\thanks{E-mail: amorandi@purdue.edu}, Daisuke Nagai${}^{2,3,4}$, Wei Cui${}^1$\\
$^{1}$ Department of Physics, Purdue University, West Lafayette, IN 47907, USA\\
$^{2}$ Department of Physics, Yale University, New Haven, CT 06520, USA\\
$^{3}$ Department of Astronomy, Yale University, New Haven, CT 06520, USA\\
$^{4}$ Yale Center for Astronomy \& Astrophysics, Yale University, New Haven, CT 06520, USA
}

\title[3D cluster parameters via the SZ effect]
{Reconstructing three-dimensional parameters of galaxy clusters via multifrequency SZ observations}
\maketitle

\begin{abstract}
The Sunyaev-Zeldovich (SZ) effect is a promising tool to study physical properties of the hot X-ray emitting intracluster medium (ICM) in galaxy clusters. To date, most SZ observations have been interpreted in combination with X-ray follow-up measurements in order to determine the ICM temperature and estimate the cluster mass. Future high-resolution, multifrequency SZ observations promise to enable detailed studies of the ICM structures, by measuring the ICM temperature through the temperature-dependent relativistic corrections. In this work we develop a non-parametric method to derive three-dimensional physical quantities, including temperature, pressure, total mass, and peculiar velocities, of galaxy clusters from SZ observations alone. We test the performance of this method using hydrodynamical simulations of galaxy clusters, in order to assess systematic uncertainties in the reconstructed physical parameters. In particular, we analyze mock Cerro Chajnantor Atacama Telescope (CCAT) SZ observations, taking into account various sources of systematic uncertainties associated with instrumental effects and astrophysical foregrounds. We show that our method enables accurate reconstruction of the three-dimensional ICM profiles, while retaining full information about the gas distribution. We discuss the application of this technique for ongoing and future multifrequency SZ observations. 
\end{abstract}


\section{Introduction}\label{intro}

Clusters of galaxies are the largest gravitationally bound structures in the universe, promising to serve as powerful cosmological probes. In the current hierarchical structure formation paradigm, galaxy clusters form out of the gravitational collapse of large peaks in the primordial density fluctuations in the early universe and grow through mass accretion. Since their evolution traces the growth of the linear density perturbations, galaxy clusters are powerful laboratories for studying the nature of dark matter and dark energy \citep[e.g.,][]{allen2008,vikhlinin2009,mantz2010}. 

Galaxy clusters can be detected through X-ray emissions from the hot ($0.5 \la T \la 10$ keV) intracluster medium (ICM). In recent years there has been significant advances in discoveries of galaxy clusters via the Sunyaev-Zeldovich (SZ) effect \citep{sunyaev1970} -- a spectral distortion of the cosmic microwave background (CMB) due to the inverse Compton scattering of the CMB photons off by the hot X-ray emitting plasma in clusters. The SZ effect signal is independent of redshift, making it an excellent method for finding clusters in the high redshift universe \citep{carlstrom2002}. Moreover, since the SZ intensity depends linearly on the density, unlike the X-ray flux, which depends on the squared density, the SZ signal is inherently sensitive to the global properties of clusters out to and beyond the virialized regions of clusters \citep{plagge2010,planck2012}. Dedicated multifrequency SZ observations, such as ACT~\footnote{{http://www.phy.princeton.edu/act}}, Planck~\footnote{{http://www.rssd.esa.int/index.php?project=planck}}, and SPT~\footnote{{http://pole.uchicago.edu}}, are discovering a large sample of high-redshift clusters \citep{vanderlinde2010,marriage2011,planck2011}, promising to provide new insights into the cluster gas physics and cosmology. 

At the same time, a number of high-resolution SZ experiments, including ALMA\footnote{Atacama Large Millimeter/submillimeter Array}, CARMA\footnote{Combined Array for Research in Millimeter-wave Astronomy}, CCAT\footnote{Cerro Chajnantor Atacama Telescope}, and MUSTANG\footnote{MUltiplexed Squid TES Array at Ninety GHz}, are also underway or planned, promising a dramatic increase in sensitivities, spatial resolution, and spectral coverage. Current high-resolution SZ observations of individual clusters have already revealed rich phenomena (e.g., shocks, substructures, relativistic particles) in the hot gaseous atmospheres of merging clusters \citep[e.g.,][]{komatsu2001,mason2010,colafrancesco2011,korngut2011,ElGordo2012,mroczkowski2012,zemcov2012,prokhorov2012}.

One of the fundamental limitations of the current SZ effect observations lies in their inability to measure the ICM temperature, independently of the X-ray measurements. Knowledge of the clusters temperatures is especially critical for estimating the cluster mass by assuming the hydrostatic equilibrium of gas in the gravitational potential. Cosmological constraints from the SZ surveys thus require extensive follow-up observations with {\it Chandra} and XMM-{\it Newton} X-ray satellites \citep[e.g.,][]{reese2002,laroque2006,bonamente2006b,benson2011}. 
 
Over the next years, high-resolution, high-sensitivity SZ observations will enable detailed studies of the thermodynamics structure of the ICM and mass modeling using the SZ data alone. The idea is to use the temperature-dependent relativistic corrections \citep{wright1979,sazonov1998,itoh1998,chluba2012} that are significant for a hot ($T_X \gesssim$~3 keV) galaxy clusters. The relativistic corrections introduce additional frequency-dependent signals, making it possible to derive the gas temperature from multifrequency SZ observations, independently of X-ray data \citep{pointecouteau1998,hansen2002,hansen2004,prokhorov2011}. Upcoming SZ experiments should further enable measurements of the peculiar velocity of the cluster and internal bulk, turbulent, and rotational gas motions within clusters \citep[e.g.,][]{chluba2002,nagai2003,sunyaev2003,diego2003}.

A variety of techniques have been developed to derive three-dimensional (3D) profile of physical quantities from two-dimensional physical observables on the plane of the sky. These techniques can be divided into two broad classes that will be denoted here as inverse method and forward-fitting method. The forward-fitting approach assumes a priori parameterized models for the 3D physical quantity in order to compare its projection with the observables in the plane of the sky \citep{mroczkowski2009,bulbul2010}. \citet{colafrancesco2010}, for example, have developed a formalism to derive the 3D ICM temperature profile from spectroscopic SZ observations alone using a parametric model. However, one of the downsides of the forward-fitting method is a need for an a priori parameterization, which might not wholly reflect the 3D physical quantities and not adequately retain the full information e.g. of the temperature profile in the gas distribution. Furthermore, the desired physical parameters of clusters extracted in this way can be prone to large systematic uncertainties, due to simplified modeling of the properties of the hot ICM and cluster morphology. The inverse method, on the other hand, introduces one or more arbitrary parameters to guarantee a smooth and physically acceptable 3D physical quantity by combining X-ray and SZ imaging data \citep{yoshikawa1999,ameglio2007}.

The main goal of our work is therefore to develop a robust non-parametric approach to derive 3D physical parameters of galaxy clusters from multifrequency SZ observations. In order to assess how accurately one can recover the 3D ICM profiles, we analyze mock CCAT observations of simulated galaxy clusters extracted from hydrodynamical simulations, taking into account various instrumental noise and astrophysical contaminations. We also briefly discuss the application of our technique for upcoming multifrequency SZ observations. Throughout this work we assume the flat $\Lambda$CDM model, with matter density parameter $\Omega_{m}=0.3$, cosmological constant density parameter $\Omega_\Lambda=0.7$, and Hubble constant $H_{0}=100h \,{\rm km\; s^{-1}\; Mpc^{-1}}$ where $h=0.7$. Unless otherwise stated, quoted errors are at the 68.3\% confidence level.

\section{The Sunyaev-Zeldovich effect}\label{szeff44}
The SZ effect is a small distortion of the spectra of the CMB, due to the inverse Compton scattering of the CMB photons off of hot ($\sim 10^7-10^8$ K) electrons of the ICM trapped into the gravitational potential well of the dark matter halo \citep{sunyaev1970,birkinshaw1999}. The CMB photons have a low probability $\tau\sim 0.01$ of interaction ($\tau$ is defined as the optical depth) with the hot electrons of the ICM, which increases on average their energy approximately by a factor $\Theta=k_{\rm {B}} T / m_e c^2$, producing a distortion of the black-body spectrum of the CMB: this appears as a decrement of the monochromatic flux of the CMB at frequencies smaller than 218 GHz and as an increment at frequencies larger than 218 GHz.

The SZ effect is expressed as a small variation in the brightness intensity $\Delta I(\nu)$ of the CMB as a function of the observation frequency:
\begin{eqnarray}
\frac{\Delta I(\nu)}{I_0} = \frac{\sigma_{T}}{m_e c^2} \int P({\bf r})\, \Psi(\nu;T({\bf r}),\beta({\bf r}))  \, dz 
\label{eq:deltai}
\end{eqnarray}
where $\sigma_T$ is the Thomson cross-section, $P({\bf r}) \equiv n_e({\bf r}) k_b T({\bf r})$ is the electron pressure of the ICM at the volume element of coordinate {\bf r}, $k_b$ is the Boltzmann constant, $z$ is the line of the sight distance, $I_0 = 2 (k_bT_{\rm cmb})^3/(hc)^2$, $T_{\rm cmb} = 2.725$ K, and $\beta = v_p/c$, where $v_p$ is the streaming velocity of the IC gas along the line of sight and it is positive (negative) for a receding (approaching) cluster.

$\Psi(\nu;T({\bf r}),\beta({\bf r}))$ takes into account the spectral shape of the SZ effect and it reads:
\begin{equation}
\Psi(\nu;T({\bf r}),\beta({\bf r})) = g(\nu;T({\bf r})) - \frac{\beta({\bf r})}{\Theta({\bf r})} h(\nu;T({\bf r}),\beta({\bf r}))\ .
\label{eq:deltai2}
\end{equation}
The first term in Equation (\ref{eq:deltai2}) is the thermal SZ (tSZ) distortion with spectral shape:
\begin{equation}
g{(\nu;T)} = \frac{x^4 e^x}{(e^x-1)^2} \left ( x \frac{e^x + 1}{e^x - 1} - 4 \right ) \left ( 1 + o_g(x; T) \right )\ ,
\label{eq:gx5}
\end{equation}
where $x = h \nu / k T_{\rm cmb}$ accounts for the frequency dependence of the SZ effect, and for the relativistic corrections related to the term $o_g(x, T)$ \citep{itoh1998}. Its magnitude is given by the Comptonization parameter: 
\begin{equation}
y \equiv \frac{\sigma_{T}}{m_e c^2} \int P({\bf r}) \, dz \;\;,
\label{proj11}
\end{equation}
If we assume a simplified isothermal parameterization of the ICM we have $y = \tau kT/(mc^2)$, where $\tau = \sigma_{\rm T} \int n_e\, dz$.

Note that the relativistic correction term $o_g(x, T)$ in Equation (\ref{eq:gx5}) becomes significant at high frequency and/or high ICM temperature: for a hot $T_X\gesssim 7$~keV clusters, the relativistic corrections are $\sim 5$\% of the total signal at 100 GHz and $\sim 30$\% at 675 GHz, making the SZ brightness very prone to the ICM temperature. The second term in Equation (\ref{eq:deltai2}) is the kinetic SZ (kSZ) distortion, which arises from the Doppler shift due to bulk motion of the electrons with respect to the rest frame of the CMB photons. The spectral shape of the kSZ effect is given by,
\begin{equation}
h(\nu;T,\beta) = \frac{x^4 \,e^x}{(e^x -1)^2}\left ( 1 + o_h(x; T,\beta) \right )\, ,
\label{eq:gx6}
\end{equation}
where the term $o_h(x; T,\beta)$ is the relativistic corrections of the kSZ effect \citep{itoh1998}, and its magnitude is proportional to $\beta/\Theta$. 

It is worth noting that in Equation (\ref{eq:deltai2}) the kinetic distortion is maximal at the crossover frequency ($\nu_0=218$~GHz), where the tSZ effect vanishes, while at lower or higher frequencies it is smaller than the thermal distortion of a factor $\sim \beta/\Theta \sim 10^{-2}-10^{-1}$ for typical values of gas temperature and peculiar velocity relevant for clusters. Note also that, in order to calculate $g{(\nu;T)}$, an a priori measurement of the gas temperature e.g., from X-ray is needed. X-ray temperature is also essential for estimating the cluster mass through the hydrostatic equilibrium assumption. For these reasons, SZ studies of cluster have have been mostly done in combination with X-ray measurements to date \citep[e.g.][]{laroque2006}.

We observe that, given the spectral signature of $\Psi(\nu;T({\bf r}),\beta({\bf r}))$, in principle it is possible to separate the physical variables $P({\bf r}),T({\bf r}), \beta({\bf r})$ by deprojecting $\Delta I(\nu)$ in Equation (\ref{eq:deltai}), in order to infer $P({\bf r}) \, \Psi(\nu;T({\bf r}),\beta({\bf r}))$. Sensitive multifrequency SZ observations can measure $\Delta I(\nu)$, thereby enabling measurements of the desired physical parameters, including $P({\bf r}),T({\bf r}), \beta({\bf r})$, independently of the X-ray measurements.

Specifically, this reconstruction is accomplished by comparing the theoretical prediction of $P({\bf r}) \, \Psi(\nu;T({\bf r}),\beta({\bf r}))$ to the deprojected observed quantities. In the next section we present a robust non-parametric approach to reconstruct 3D physical quantities from two-dimensional projected SZ maps by inverting Equation (\ref{eq:deltai}).

\section{Spectral deprojection technique}\label{apecdepte}
A variety of techniques have been developed to derive 3D physical quantities (e.g. $P({\bf r}) \, \Psi(\nu;T({\bf r}),\beta({\bf r}))$) from two-dimensional physical observables on the plane of the sky (e.g. $\Delta I(\nu)$). As discussed in \S~\ref{intro}, in this work, we develop a robust non-parametric approach for inverting Equation (\ref{eq:deltai}) and constraining 3D temperature and pressure profiles from SZ-only multifrequency observations.

We assume that the cluster is spherically symmetric and it has a onion--like structure with $n$ concentric spherical shells, each characterized by uniform gas density and temperature within it. Therefore, the cluster image in projection is divided into $m$ rings (or annuli) of area ${\bf A}=(A_1,A_2,...,A_m)$, which are assumed to have the same radii of the 3D spherical shells. The SZ boundary (the most external ring) will be represented by the $m$th ring of area $A_m$ and radius $R_m$. Let us define $\epsilon_i$ as the SZ signal to be recovered from the deprojection method within the $i$th shell. In our analysis {\textbf{s}} and {\textbf{$\epsilon$}} read:
\begin{equation}
{\bf s}={\Delta I(\nu)}/{I_0}\, , \quad {\bf \epsilon}=\frac{\sigma_{T}}{m_e c^2}\, P({\bf r}) \, \Psi(\nu;T({\bf r}),\beta({\bf r})). 
\label{kk0}
\end{equation}
with ${\bf r}=R_i,\; i=1,...,m$. In this way, the contribution of the $i$-th shell to the SZ surface brightness in the ring $j$ of the image will be given by 
\begin{equation}
{\bf s }=\left({\bf \sf V}\#{\bf \epsilon}\right)/{\bf A}
\label{kk}
\end{equation}
where the operator $\#$ indicates the matrix product (rows by columns). The matrix ${\bf \sf V_i^j}$ is a rectangular $m\times n$ matrix, with the $m$-dimensional column vectors ${\bf  V^1,V^2, ...,V^n}$ representing the effective volumes, i.e. the volume of the $j$-th shell contained inside the $i$-th annulus (with $j \ge i$) and corrected by the gradient of $\epsilon$ inside the $j$-th shell \citep[see Appendix~B of][]{morandi2007a}. We assumed that $n>m$: this means that in $j$th ring of the image we account also for the so-called edge effect, i.e. for the contribution of the shells $(m+1,m+1,..,n)$ of radius $R>R_{\rm bound}$.

\begin{figure}
\begin{center}
\psfig{figure=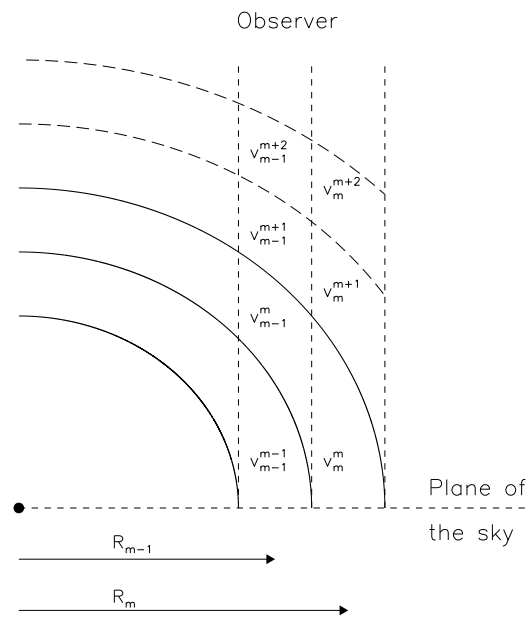,width=0.48\textwidth}
\caption{Illustration of the onion-skin model adopted for the geometrical deprojection. We assume that the cluster is spherically symmetric and it has a onion--like structure, with $n$ concentric spherical shells and $m$ rings or annuli ($n>m$). The matrix ${\bf \sf V_i^j}$ is a rectangular $m\times n$ matrix, whose entries represent the volume of the $j$-th shell contained inside the $i$-th annulus (with $j \ge i$). The SZ boundary (the most external annulus) will be represented by the $m$th ring of area $A_m$ and radius $R_m$. The shells $(m+1,m+1,..,n)$ of radius $R>R_m$ (long-dashed line) account for the edge effect.}
\label{fig1}
\end{center}
\end{figure}

Now, in order to infer ${\bf \epsilon}$, we introduce two types of information. The first is a statistical information and allows us to obtain a solution which reproduces the data with a known accuracy. In our case, since the error on the data are assumed to follow a normal distribution, the $\chi^2$ gives the accuracy of the solution:
\begin{equation}
\chi^2={[{\bf s}-\left({\bf \sf V}\#{\bf  \epsilon}\right)/{\bf A}]}^{\rm t}\#{\bf \sf C^{-1}}\# {[{\bf s}-\left({\bf \sf V}\#{\bf \epsilon}\right)/{\bf A}]}\ ,
\label{kk2}
\end{equation}
where ${\bf \sf C^{-1}}$ is the error covariance matrix. Nevertheless, it is well known that this is an ill-posed problem, meaning that the solution of Equation (\ref{kk2}) is unstable with respect to the errors in the data, i.e. a small error in the initial data can result in much larger errors in the desired solution.

The second information is related to the physics, and it allows us to ensure that the solution is physically acceptable. We use an approach known as regularization, which involves introducing additional a priori information in order to solve an ill-posed problem. This information is usually of the form of a penalty for complexity, such as restrictions for smoothness, by reducing the number of degrees of freedom of the problem. It includes various facts known about the solution a priori, e.g. the total cumulative mass in clusters must be positive, and monotonically increasing regardless the observational data.

It is then necessary to introduce a means of balancing the regularization constraints with the accuracy to which the solution reproduces the data, which is parameterized by $\chi^2$. This has been accomplished by using the Lagrangian multipliers to find a solution; in particular we minimize the function:
\begin{equation}
L(\epsilon, \lambda) = \chi^2(\epsilon) + \lambda C(\epsilon)
\label{kk3}
\end{equation}
where $C(\epsilon)$ is a function that is minimized when the solution best satisfies the regularity constraint, and $\lambda$ is the smoothing parameter.

We adopt the Philips-Towmey regularization method \citep{phillips1962,twomey1963}. Generally, this method consists of minimizing the sum of the squares of the $k^{th}$ order derivates of the solution around each data point. For our purposes, we define the ``smoothness'' constraint as the minimum of the sum of the squares of the second derivatives, so that:
\begin{equation}
C(f) = \sum_{j=2}^{n-1} (a_j \epsilon_{j-1} + 2b_j*\epsilon_{j} + c_j\epsilon_{j+1})^2=\epsilon\#{\bf \sf H}\#\epsilon
\label{kk4}
\end{equation}
where the $a_j=1$, $b_j=-1$, $c_j=1$, and the matrix ${\bf \sf H}$ is defined in \cite{twomey1963}.

The solution of Equation (\ref{kk3}) reads \citep[see][]{bouchet1995}:
\begin{equation}
\epsilon={\left({{\bf \sf V}^t}\#{\bf \sf C^{-1}}\#{\bf \sf V}+\lambda\, H\right)}^{-1} \#{\bf \sf V}^t\#{\bf \sf C^{-1}}\#\left({\bf s}\cdot {\bf A}\right)
\label{kksd}
\end{equation}

By varying $\lambda$, we can therefore vary the degree to which the solution is dominated by consistency with the data (i.e. it is statistically acceptable) or with the regularizing constraints (i.e. it is physically acceptable). The choice of $\lambda$ is therefore critical to obtaining a reliable solution. To choose the value of $\lambda$ we use the cross-validation, which offers an objective and robust way of choosing the smoothing parameter. Cross-validation consists of predicting each data point ${\bf s}$ by finding a solution using all the data but that point. The best value of $\lambda$ is the one that predicts the observations best, i.e. the one which minimizes the average quadratic errors between the data ${\bf s}$ and their predicted values or the reduced $\chi^2$ \cite[see e.g.,][for further details]{bouchet1995}.

It is worth emphasizing the improvement of this method with respect to the standard deprojection technique in X-ray spectra (or surface brightness), where customarily the desired solution is inferred via inversion of Equation (\ref{kk}) under the assumption that $n=m$, neglecting the edge effect. In this case ${\bf \sf V}$ would be an upper triangular matrix, whose inversion would allow us to determine ${\bf \epsilon}$ via Equation (\ref{kk}). While this approach is commonly employed in the deprojection of X-ray physical observables, such as the X-ray temperature or brightness \citep{1983ApJ...272..439K,morandi2007a}, for SZ data it would not be satisfactory, because the contribution of the shells with radius greater than the SZ boundary is non-negligible (since the SZ effect is linearly proportional to the electronic density, unlike the X-ray brightness), and the measurement errors ${\bf \sigma}$ are not taken into account in the determination of the expectation value of ${\bf \epsilon}$. Alternatively, one could extrapolate the profile of ${\bf s}$ via an analytical function out to $R_n>R_m$ and consider an $n\times n$ upper triangular matrix ${\bf \sf V}$ in order to infer ${\bf \epsilon}$ via Equation (\ref{kk}): nevertheless, in this case the solution could be systematically biased, depending on the assumed extrapolated profile and on the choice of outermost bin. Note that in our approach we corrected for the edge effect without assuming any a priori profile of ${\bf s}$ beyond the SZ boundary.

\begin{table}
\begin{center}
\caption[]{Specifications of the CCAT SZ cluster observations. The stated frequency refers to the central frequency of the bandwidth; the noise level accounts for the confusion limit by dusty star-forming galaxies and refers to an integration time of 16 hours.}
\begin{tabular}{l@{\hspace{1.3em}} c@{\hspace{1.3em}} c@{\hspace{1.3em}} }
\hline \
$\nu$   & beam & rms \\
 (GHz)       &     (arcsec)     &     ($\mu$Jy/beam)  \\
\hline   \\
 93   &    27.0     &  20.0  \\
148   &    16.8     &  31.2  \\
221   &    10.8     &  25.0  \\
272   &     9.0     &  26.2  \\
347   &     7.2     &  38.8  \\
406   &     6.0     &  75.0  \\
\hline \\ 
\end{tabular}
\label{tabdon2}
\end{center}
\end{table}

\section{Mock CCAT Simulations}\label{apecdepte2}

In this work we analyze mock CCAT simulations of galaxy clusters. We start by describing the CCAT telescope in \S \ref{ccat} and the simulated cluster sample in \S \ref{dmicm2}. We then describe how we generate SZ maps from these simulations (\S \ref{apecdepte3}), including the impact of relativistic electrons (\S \ref{cbianv}) and the primary CMB anisotropies (\S \ref{cbian}). 

\subsection{CCAT SZ Observations}\label{ccat}

Cerro Chajnantor Atacama Telescope (CCAT) is 25-meter submillimeter telescope that will be located at 5600 m altitude on Cerro Chajnantor in the Andes mountains of northern Chile. CCAT will combine high sensitivity, a wide field of view, and a broad wavelength range to provide an unprecedented capability for deep, large area multicolor submillimeter surveys. In Table~\ref{tabdon2} we summarize the specifications of the proposed CCAT telescope, including frequency channels, angular resolution, and sensitivities for the integration time of 16 hours, which represents a feasible integration time for this facility.

\begin{table}
\begin{center}
\caption[]{Properties of the simulated cluster sample. We report $R_{500}$, the true global mass-weighted temperature $T_{\rm mw}$ of the clusters, the minor-major and the intermediate-major axis ratios of the ICM and DM. The last column refers to the centroid-shift for the $x$, $y$ and $z$ projections, respectively.}
\begin{tabular}{l@{\hspace{0.5em}} c@{\hspace{0.5em}} c@{\hspace{0.5em}} c@{\hspace{0.5em}} c@{\hspace{0.5em}} c@{\hspace{0.5em}} }
\hline \
object  & $R_{500}$ & $T_{\rm mw}$ & $\eta_1-\eta_2$ & $ \eta_1-\eta_2$   & $w$ \\
        &      (kpc) &      (keV)     &     ({\rm ICM})             &     ({\rm tot})            &    ($x/y/z$)           \\
\hline   \\
CL101   &   1657 & 7.44 &  0.95$-$0.98  &  0.79$-$0.88  &  0.011$/$0.016$/$0.012   \\
CL102   &   1397 & 5.63 &  0.92$-$0.93  &  0.85$-$0.95  &  0.023$/$0.023$/$0.009   \\
CL103   &   1420 & 4.84 &  0.91$-$0.99  &  0.83$-$0.92  &  0.017$/$0.050$/$0.055   \\
CL104   &   1394 & 6.61 &  0.90$-$0.95  &  0.79$-$0.91  &  0.002$/$0.005$/$0.004   \\
CL105   &   1347 & 5.67 &  0.95$-$0.96  &  0.80$-$0.89  &  0.015$/$0.005$/$0.013   \\
CL106   &   1202 & 4.54 &  0.72$-$0.85  &  0.67$-$0.88  &  0.039$/$0.053$/$0.037   \\
CL107   &   1089 & 3.61 &  0.76$-$0.80  &  0.72$-$0.75  &  0.029$/$0.022$/$0.034    \\
CL3     &   1016 & 3.37 &  0.80$-$0.88  &  0.73$-$0.85  &  0.007$/$0.010$/$0.007   \\
\hline \\ 
\end{tabular}
\label{tabdon}
\end{center}
\end{table}

\subsection{Simulated Clusters Sample}\label{dmicm2}
We analyzed a sample of high-resolution hydrodynamical simulations of galaxy clusters formation from \citep[][hereafter N07a,b]{nagai2007a,nagai2007b}, which were performed using the ART code \citep{kravtsov2002,rudd2008}. In the present work, we analyze the z=0 outputs of the simulations that include radiative cooling, star formation, metal enrichment and stellar feedback at low redshift. In particular, we focused on a subsample of 8 clusters with the global mass-weighted temperature $T_{\rm mw} \gesssim 3$~keV, where the relativistic corrections would be large enough to robustly measure the cluster gas temperature from multifrequency SZ-only data. Each cluster is simulated using a $128^3$ uniform grid with 8 levels of refinement. Clusters are selected from 120$h^{-1}$~Mpc computational boxes, achieving peak spatial resolution of $\approx 3.6h^{-1}$~kpc. The dark matter particle mass in the region surrounding the cluster is $9\times 10^8 h^{-1}M_{\odot}$, while the rest of the simulation volume is followed with lower mass and spatial resolution. We refer readers to N07 for the details of these simulations. 

In order to assess the effect of cluster shapes on the derived 3D profiles, we calculated the intermediate-major and minor-major axis ratios of both the DM and the ICM. Assuming an onion--like density distribution in concentric ellipsoids, it is possible to determine the axial ratios by diagonalizing the inertia tensor \citep{splinter1997}:
\begin{equation}
U_{\rm ij} = \sum {x_{\rm i}x_{\rm j}}\; \rho;\quad (i,j=1,2,3),
\end{equation}
where the sum is over all points $x_1= x$, $x_2= y$, $x_3= z$ of density $\rho$. With this definition we have the minor-major and intermediate-major axis ratios $\eta_1$ and $\eta_2$, respectively:
\begin{equation}
\eta_1 = \left( \frac{U_{\rm zz}}{U_{\rm xx}} \right)^{1/2}{\rm and}\;\;\; \,\eta_2 = \left( \frac{U_{\rm yy}}{U_{\rm xx}} \right)^{1/2} \;\;\; 
\end{equation} 
with $\eta_1 \leqslant \eta_2  \leqslant 1$, and where $U_{\rm xx}$, $U_{\rm yy}$, and $U_{\rm zz}$ are the principal components of the diagonalized tensor, with $U_{\rm zz}\leqslant U_{\rm  yy}\leqslant U_{\rm xx}$. We considered both the cases where $\rho=\rho_{\rm ICM}$ and $\rho=\rho_{\rm tot}$, for the ICM and DM principal axis ratios, respectively. We also point out that a few clusters (CL103, CL106 and CL3) show visible substructures with the resolution and sensitivity of the observations under consideration. The most prominent substructures have been identified by visual inspection and masked out before the analysis.

We then evaluated the SZ centroid-shift $w$ of the Compton parameter image, which is a proxy of the state of relaxation and level of substructures of a cluster. $w$ was determined in a series of circular apertures centered on the cluster SZ peak out to the SZ boundary radius, roughly corresponding to $R_{500}$\footnote{$R_{500}$ is the radius of a sphere centered in a local minimum of the potential and enclosing an average density of $\rho=500\, \rho_{cr,z}$, with $\rho_{\rm c, z}\equiv 3H(z)^2/ 8 \pi G$ being the critical density of the universe at redshift $z$, $H(z)\equiv \left[\Omega_M (1+z)^3  + \Omega_{\Lambda}\right]^{1/2}\,H_{0}$.}. $w$ was defined as the standard deviation of the projected separations between the peak and centroid in units of $R_{500}$. Table~\ref{tabdon} presents the list of objects analyzed in the present paper, including $R_{500}$, their temperature, principal axis ratios of both ICM and DM, and the measured centroid shifts.

\subsection{SZ maps}\label{apecdepte3}

\begin{figure*}
\begin{center}
 \hbox{
\psfig{figure=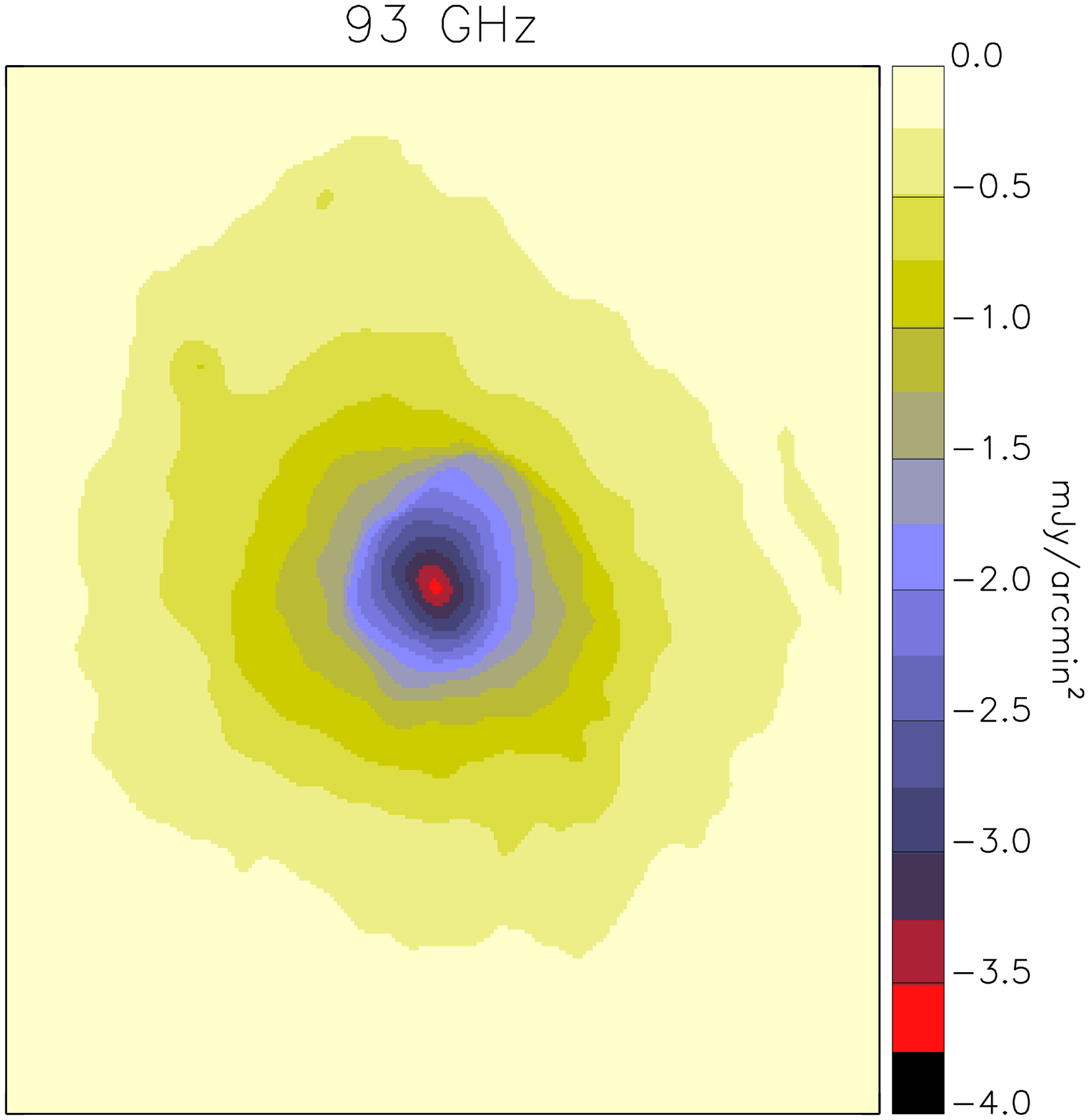,width=0.33\textwidth}
\psfig{figure=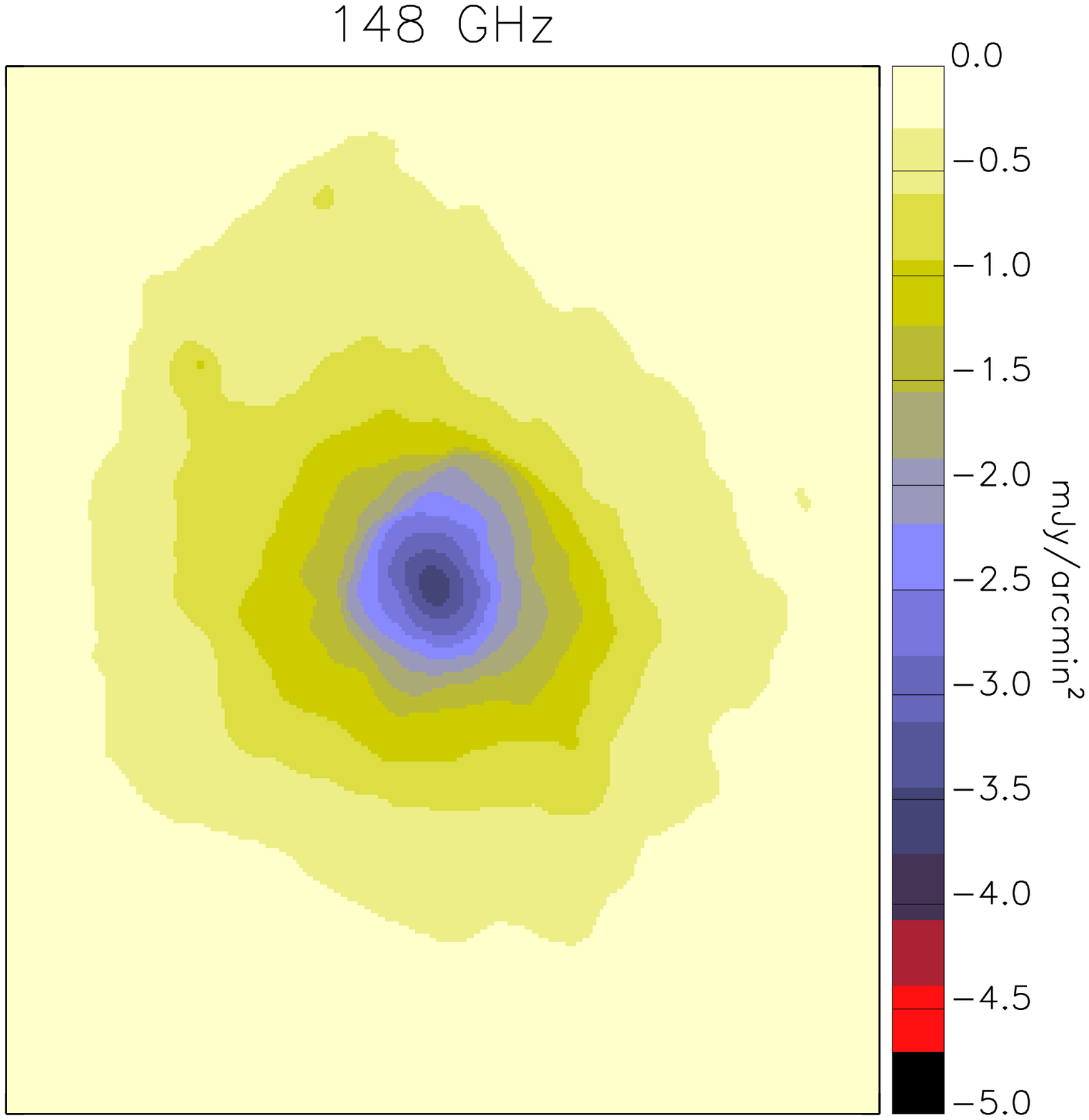,width=0.33\textwidth}
\psfig{figure=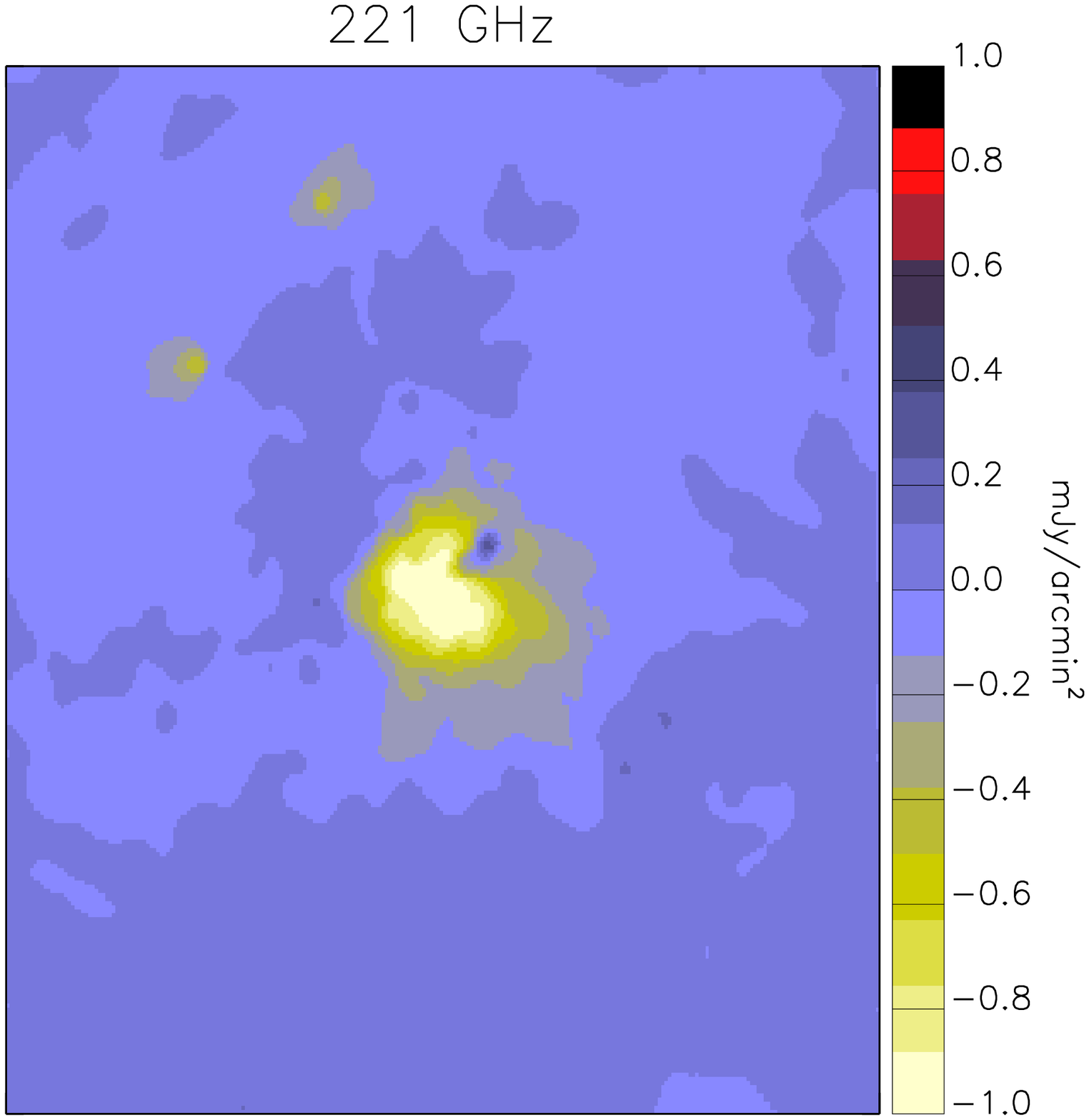,width=0.33\textwidth}
}
 \hbox{
\psfig{figure=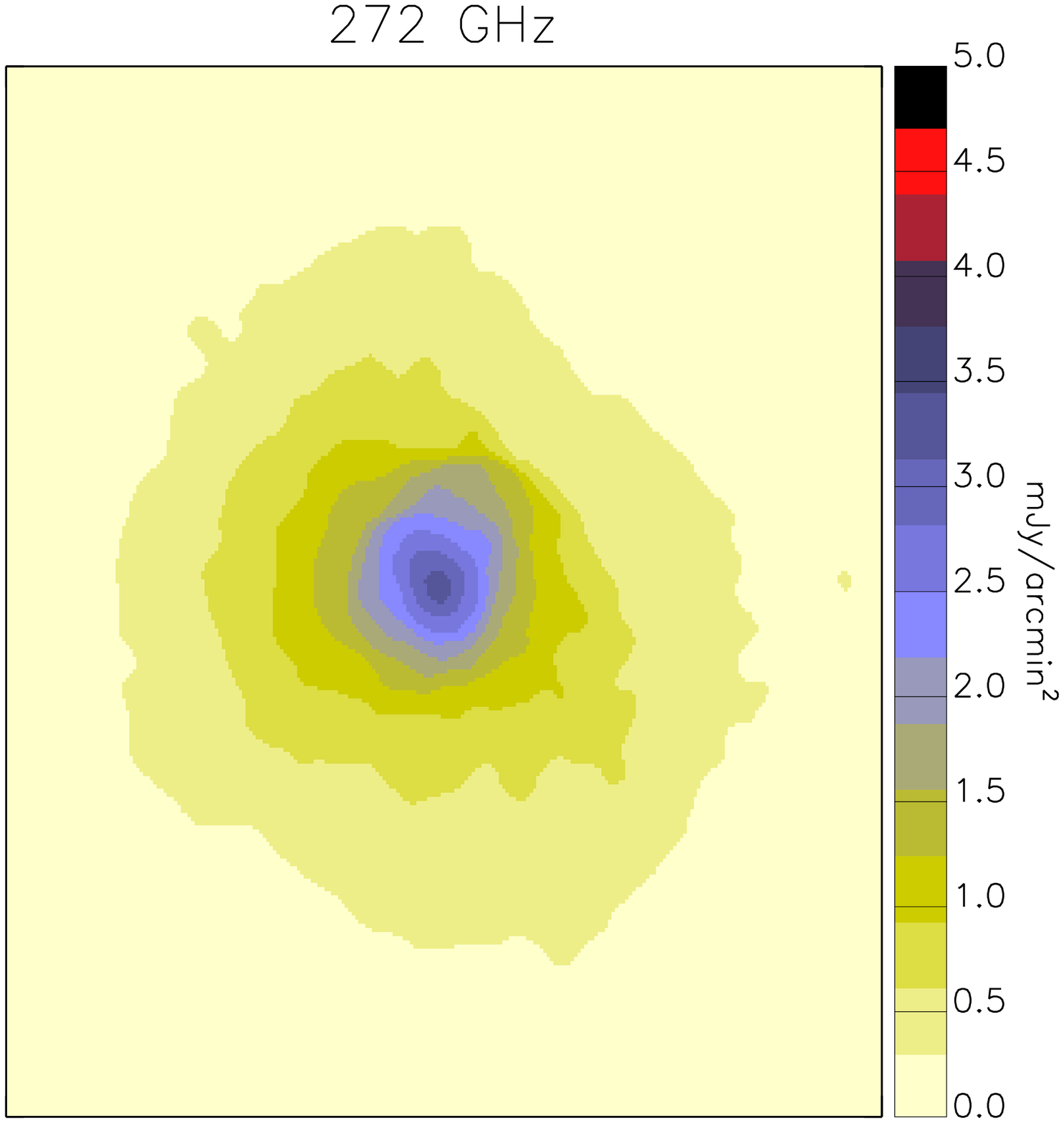,width=0.33\textwidth}
\psfig{figure=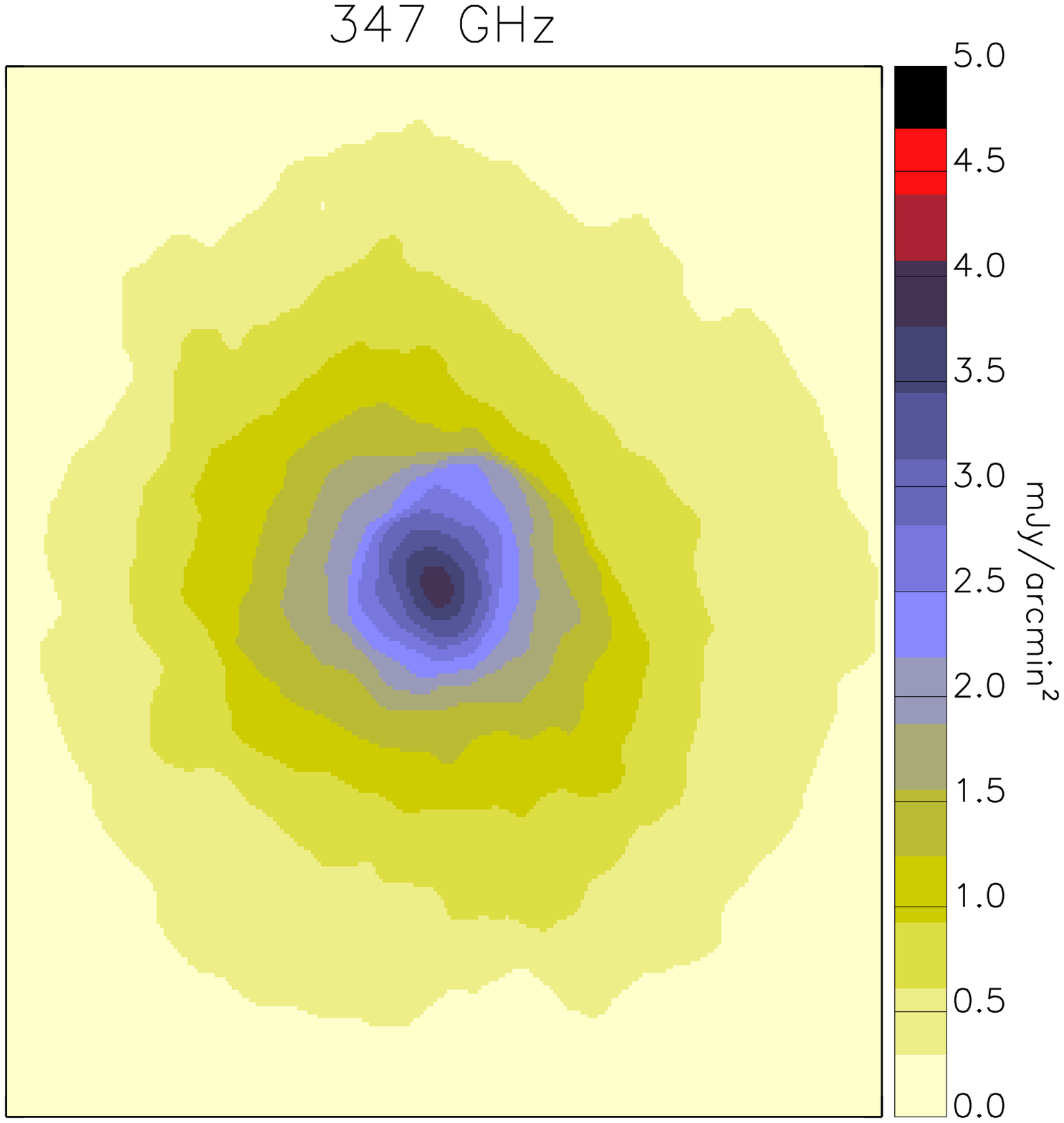,width=0.33\textwidth}
\psfig{figure=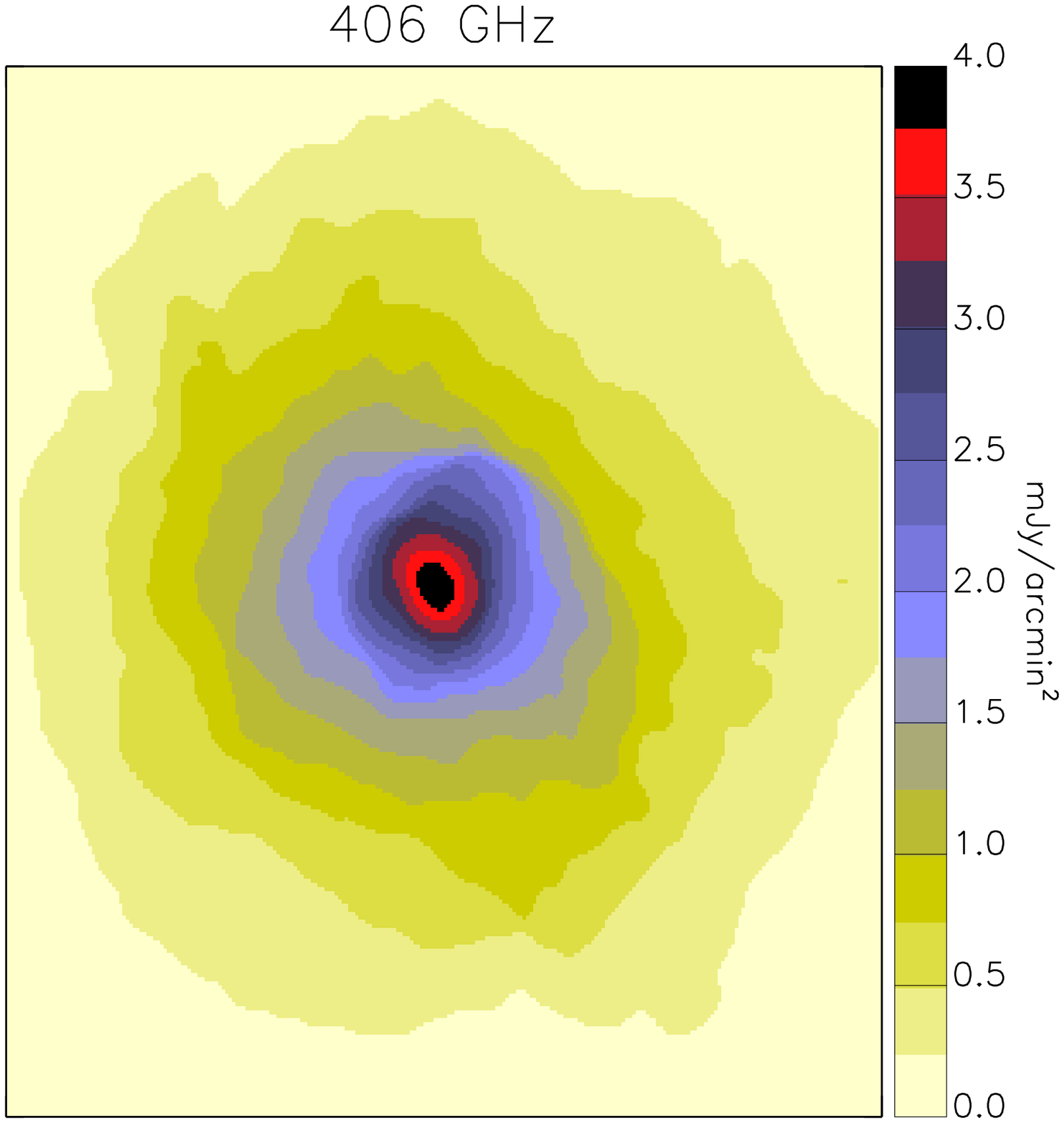,width=0.33\textwidth}
}
\caption[]{The SZ brightness $\Delta I(\nu)$ maps of the cluster CL101 viewed along the $z$ projection at 6 CCAT bands. The size of the region shown in the figure is 5.7 Mpc centered on the minimum of cluster potential, with a pixel size of 22 kpc.}
\label{entps3xkn3}
\end{center}
\end{figure*}

We generate mock CCAT observations of the simulated clusters at 6 bands between $\sim$100-800 GHz and with the beam size and expected noise level for 16 hours of CCAT observations. We used the outputs of the simulated galaxy clusters at $z=0$ and placed it at the observing redshift of $z=0.2$. Specifically, we simulated the SZ intensity maps $\Delta I(\nu)$ at each frequency by means of Equation (\ref{eq:deltai}) and by using the 3D gas pressure, temperature and peculiar velocity for the simulated galaxy clusters. For each cluster, the mock CCAT data (i.e., the SZ brightness $\Delta I(\nu)$) are created for three orthogonal projections by projecting both thermal and kinematic SZ signals along the $x$, $y$, and $z$ coordinate axes, accounting for the line-of-sight velocity structures ($v_x$, $v_y$ and $v_z$) in simulations. 

Figure \ref{entps3xkn3} presents the SZ brightness maps of CL101 (recovered by applying Equation \ref{proj11}) at each CCAT band. The size of the region shown in the figure is $\sim 5.7$~Mpc with our cosmological parameters, with a pixel size of $\sim 22$~kpc, and the region is centered on the minimum of cluster potential. We also model measurements at a given resolution by convolving the maps with a Gaussian beam profile of appropriate width corresponding to the resolution of CCAT at each frequency.

\subsection{Relativistic Electrons}
\label{cbianv}

The presence of significant energetic electron populations in many clusters has been known from measurements of diffuse radio emission and non-thermal X-ray emission from relativistic electrons in several nearby clusters. The range of electron energies implied from these measurements is roughly 1-100 GeV. This can have appreciable impact on the reconstruction of the physical properties of the ICM from the SZ data, especially on the measurements of the cluster's peculiar velocity discussed in \S~\ref{phys11d}. 

To assess the magnitude of this effect, we have included in the simulations the additional intensity change in the CMB spectrum due to non-thermal electrons: 
\begin{equation}
\frac{\Delta I_{\rm rel}(\nu)}{I_0}=-\frac{x^3}{(e^x-1)^2}\,\tau_{\rm rel}
\label{kksd4}
\end{equation}
where $\tau_{\rm rel}= \sigma_{\rm T} \int n_{\rm e,rel}\, dz$ is the optical depth of the relativistic electrons.

For simplicity, the non-thermal emission region has been assumed to have spherical symmetry with an electron spectrum:
\begin{equation}
N_e(p,r) = k_0 p^{-s_1} \cdot g_e(r) \;\;\;\;\; p\geq p_1 \; \; .
\label{eq.elent}
\end{equation}
Here we assume $s_1=3$, $p_1=10$, $k_0=3.7\times 10^{-3}$ cm$^{-3}$ and a radial distribution of the electron population given by:
\begin{equation}
g_e(r)=\left[ 1+ \left(\frac{r}{r_c}\right)^2 \right]^{-q_e}
\label{eq.elespaz}
\end{equation}
with $r_c=10$ kpc and $q_e=0.5$. With this choice of $k_0$, the non-thermal electron energy constitutes $\lesssim 1\%$ of the thermal energy in the cluster, in agreement with the observations of Coma cluster \citep{shimon2002}. 

Therefore, the contribution of the non-thermal SZ signal is generally small compared to the tSZ signal ($\Delta I_{\rm rel}/\Delta I_{\rm th} \ll 10^{-2}$) in most bands, except near the crossover frequency at 218~GHz (e.g., $\Delta I_{\rm rel}/\Delta I_{\rm th} \lesssim 20\%$ at 211~GHz). Hence relativistic electrons have a negligible impact on the cluster temperature determination \citep[see also,][]{shimon2002}, but they can lead to significant biases on the peculiar velocity determination, which stems mainly from SZ measurements around 218~GHz. Note that, even if we increase $k_0$ by a factor of five in order to mimic the properties of clusters hosting strong radio halos \citep[e.g. A2199,][]{shimon2002}, we still find a small impact of the relativistic electrons on the derived cluster temperature.

\begin{figure*}
\begin{center}
\hbox{
\psfig{figure=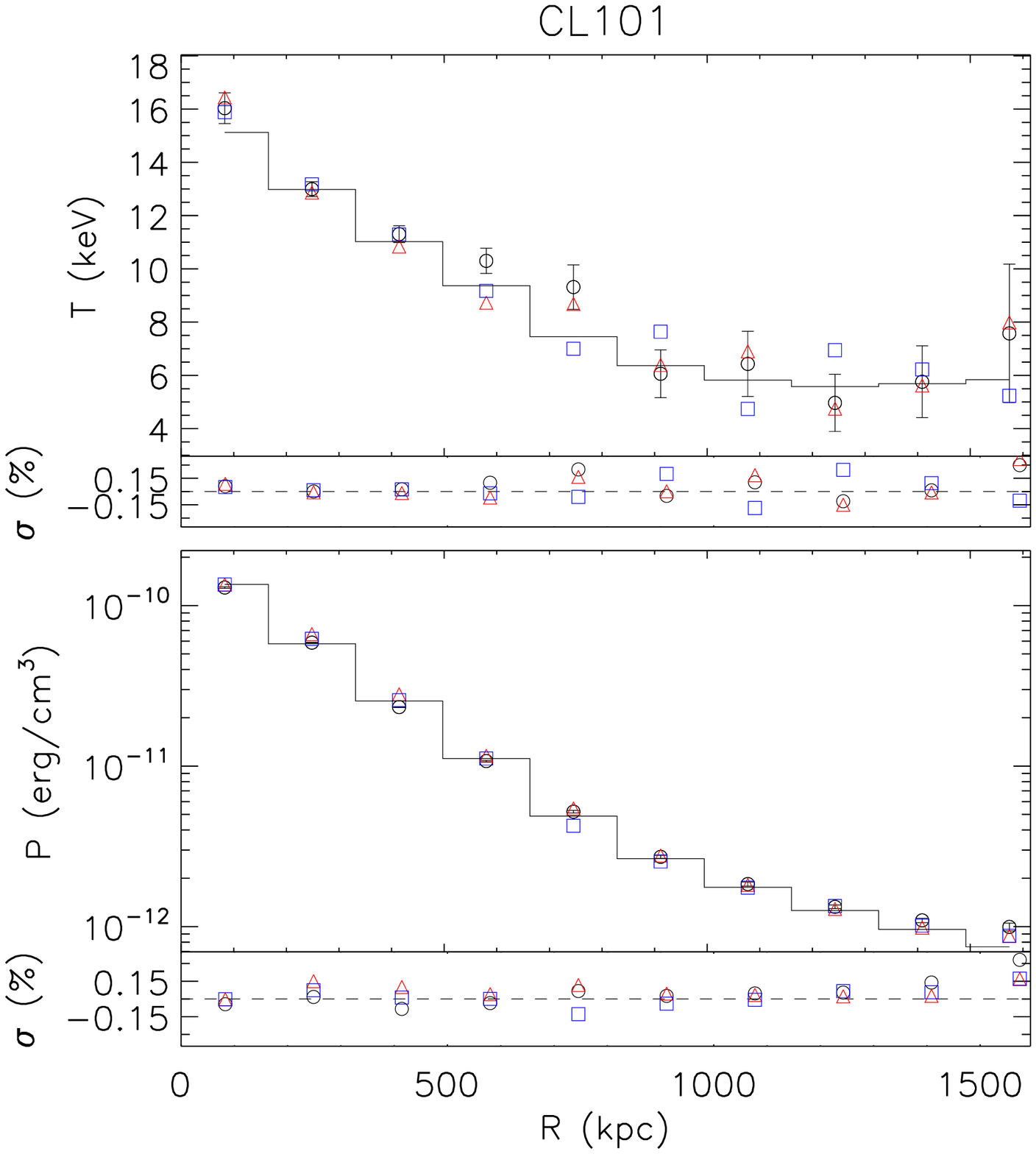,width=0.47\textwidth}
\psfig{figure=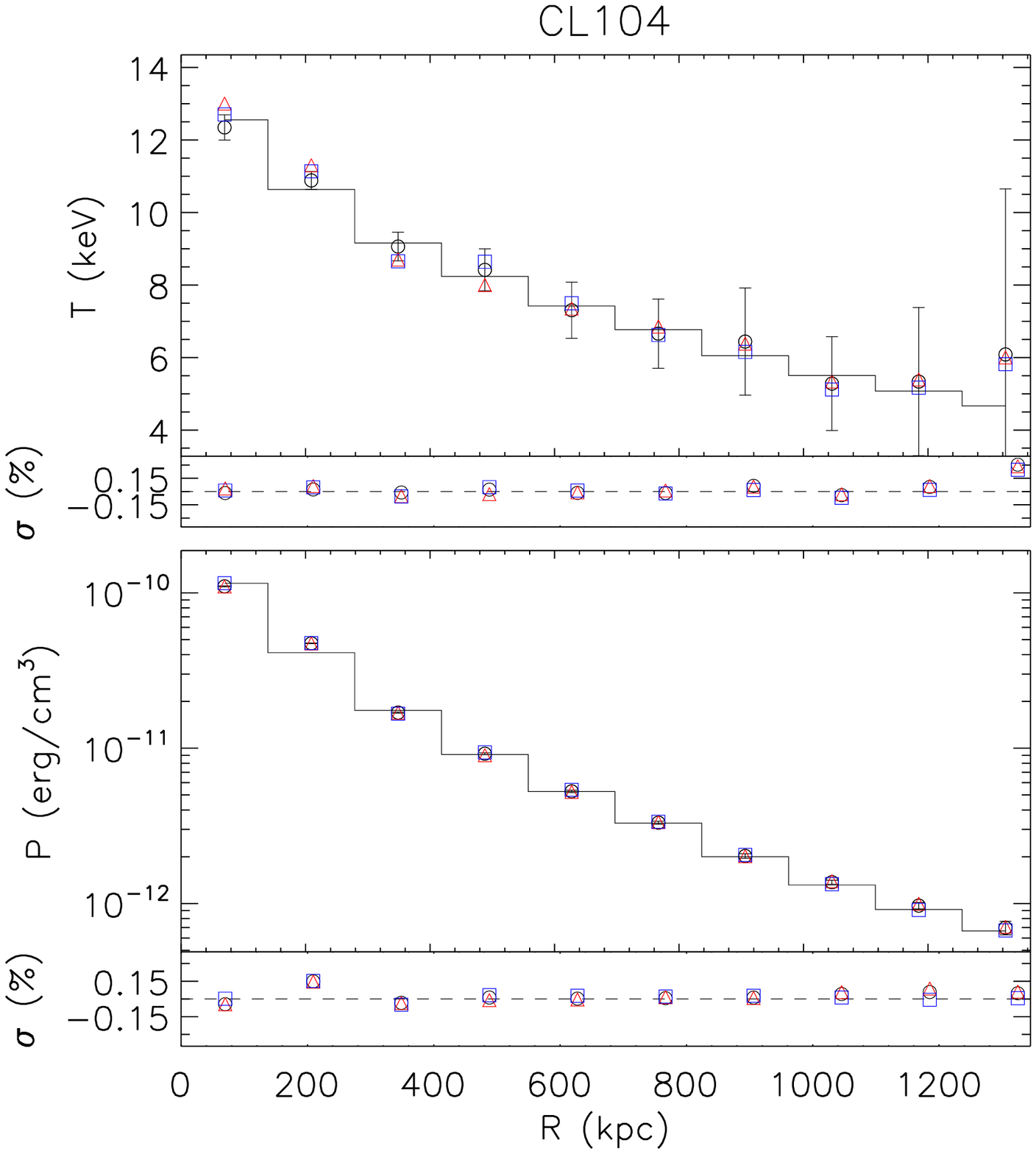,width=0.47\textwidth}
}
\caption[]{The temperature (upper panel) and pressure (lower panel) profiles retrieved from mock CCAT analyses of CL101 and CL104. The black-solid represents the true mass-weighted temperature (pressure) of the clusters in the $i$th spherical shell, whose width is graphically represented by the horizontal solid lines. The open points with errorbars are the temperature (pressure) as retrieved from the multifrequency SZ-only analysis for the $z$ projection, while the triangles (squares) refer to the $x$ ($y$) projection (errorbars are omitted for clarity). We also report the fractional total scatter at the bottom of each panel. }
\label{ent246g}
\end{center}
\end{figure*}

\subsection{Primary CMB anisotropies}
\label{cbian}
Primary CMB anisotropies are another potential source of contamination in SZ observations. CMB anisotropies, $\Delta T$, translate into uncertainties on the SZ signal ${\Delta I(\nu)}$ by the derivative of the blackbody spectrum with respect to the temperature, i.e. ${\Delta I(\nu)}/{I_0}= {x^4 e^x}/(e^x-1)^2\; \Delta T/ T_{\rm cmb}$. To assess the effect of the primary CMB anisotropies, we first produce a realistic CMB map $T_{\rm cmb}+\Delta T$ in order to generate the ${\Delta I(\nu)}/{I_0}$ simulated images described in \S \ref{apecdepte3}. We then explicitly include in the covariance matrix ${\bf \sf C}$ (see \S \ref{apecdepte}) both the uncertainties due to the CMB correlated signal and the thermal noise of the instrument as:
\begin{equation}
{\bf \sf C=N+C_{\rm cmb}},
 \label{cov33}
\end{equation}
where ${\bf \sf C_{\rm cmb}}$ is the CMB covariance matrix, ${\bf \sf N}$ is the noise diagonal matrix $\sigma_i^{-2}$, and ${\bf \sigma}$ is the thermal noise of the instrumental.

We simulated realistic maps of the CMB given an input power spectrum $C_l$ according to the WMAP 7-year results \citep{jarosik2010}. We then evaluated the product of the WMAP power spectrum $C_l$ and the beam transfer function $\sim \exp(-0.5 \, l\,(l+1)\,\theta^2)$, with $l=\pi/\arctan{[0.5\sqrt{(u^2+v^2)}]}\sim \sqrt{2\pi\, (u^2+v^2)}$ ($u$ and $v$ are the spatial frequencies), $\theta=\theta_{\rm fwhm}/\sqrt{8\log{(2)}}$, and $\theta_{\rm fwhm}$ is the beam of the telescope.

Given a CMB image $\Delta T(x,y)$ characterized by $N\times N$ pixels and angular resolution $\Delta x=\Delta y$, we performed Monte-Carlo simulations of $\Delta T(x,y)$ in the Fourier domain by creating a complex visibility $V(u,v)$ with $\Delta u= \Delta v=1/(N\, \Delta x)$. For each point in the Fourier domain we compute its distance $l$ and the corresponding $C_l$ from the input power spectrum. We set the expectation value and variance of $V(v,u)$ to be zero and $\sqrt{C_l}\Delta u$, respectively. We then perform a Gaussian randomization of both the real and imaginary parts of $V(u,v)$, taking into account its conjugate symmetry. We set the central visibility $V(0,0)$ to be zero, so the sum of the CMB fluctuations $\Delta T$ is zero over the whole image. We then obtain realization of the CMB map $\Delta T(x,y)$ by performing the inverse Fourier transform of $V(u,v)$.

In order to assess the impact of primary CMB anisotropies, we employed a lead$-$trail (L$-$T) differencing scheme where two fields (the source and the CMB fields) are observed in succession at the same hour angle and subtracted from each other. This removes the ground signal and any other potential spurious signals. The angular distance between the L$-$T fields has been assumed to be 45 arcmin. The impact of the primary CMB anisotropies depends mainly on the angular resolution (beam), the angular size of the cluster, the width of the annuli where the signal $\Psi(\nu;T({\bf r}),\beta({\bf r}))$ is averaged out and deprojected, and the angular distance between the L$-$T fields. The contribution of the primary CMB anisotropies increases toward larger angular scale out to about one degree, and it becomes a dominant source of uncertainties beyond the SZ boundary of the observation on scale $\gesssim 4-6$ arcmin, roughly corresponding to $R_{500}$. This also sets the outer radius out to which the three dimensional profiles can be reliably measured using the multifrequency SZ observations. On scales $\lesssim 1$ arcmin the CMB anisotropies is negligible compared to the thermal noise.

\begin{figure*}
\begin{center}
\hbox{
\psfig{figure=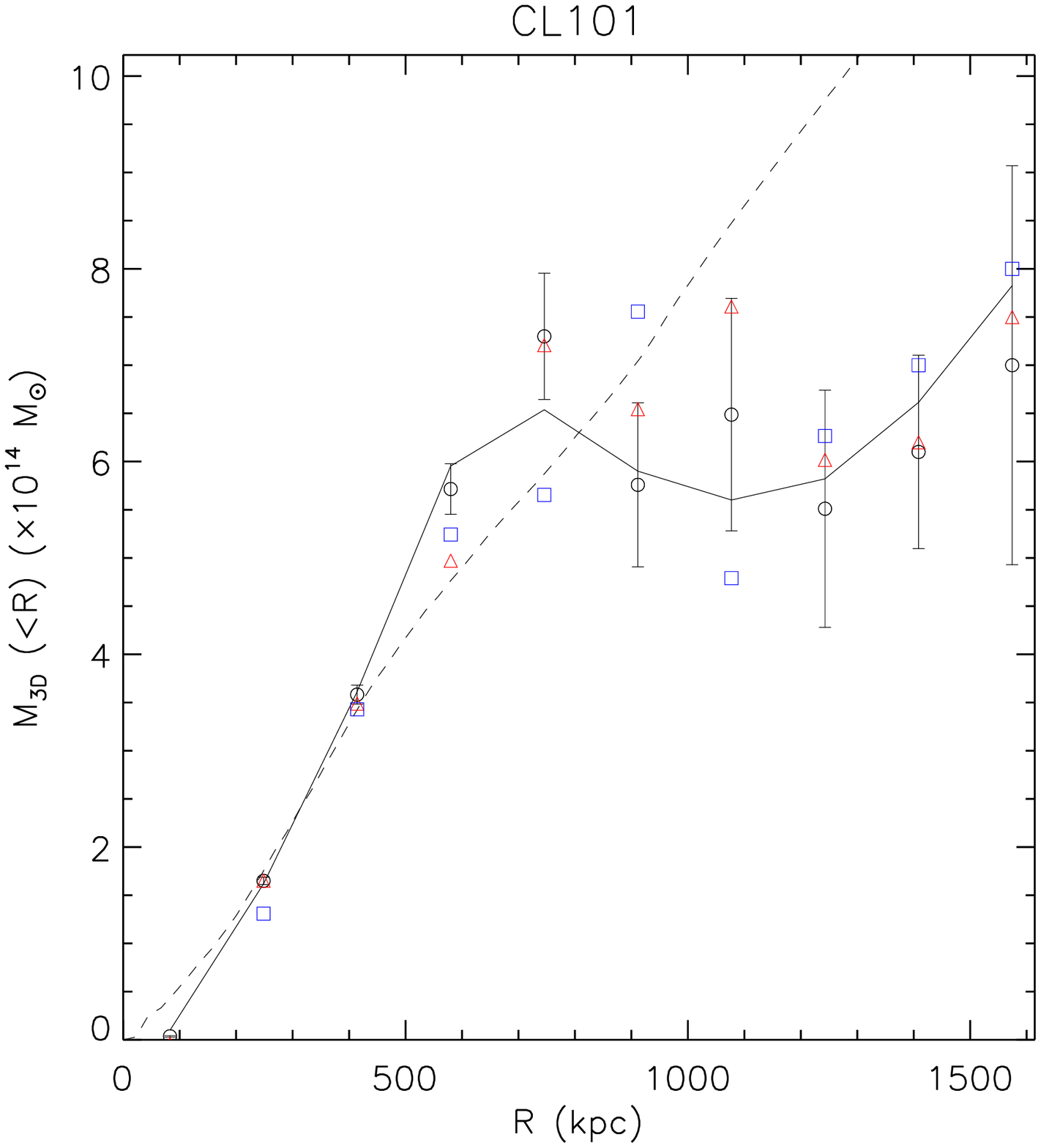,width=0.47\textwidth}
\psfig{figure=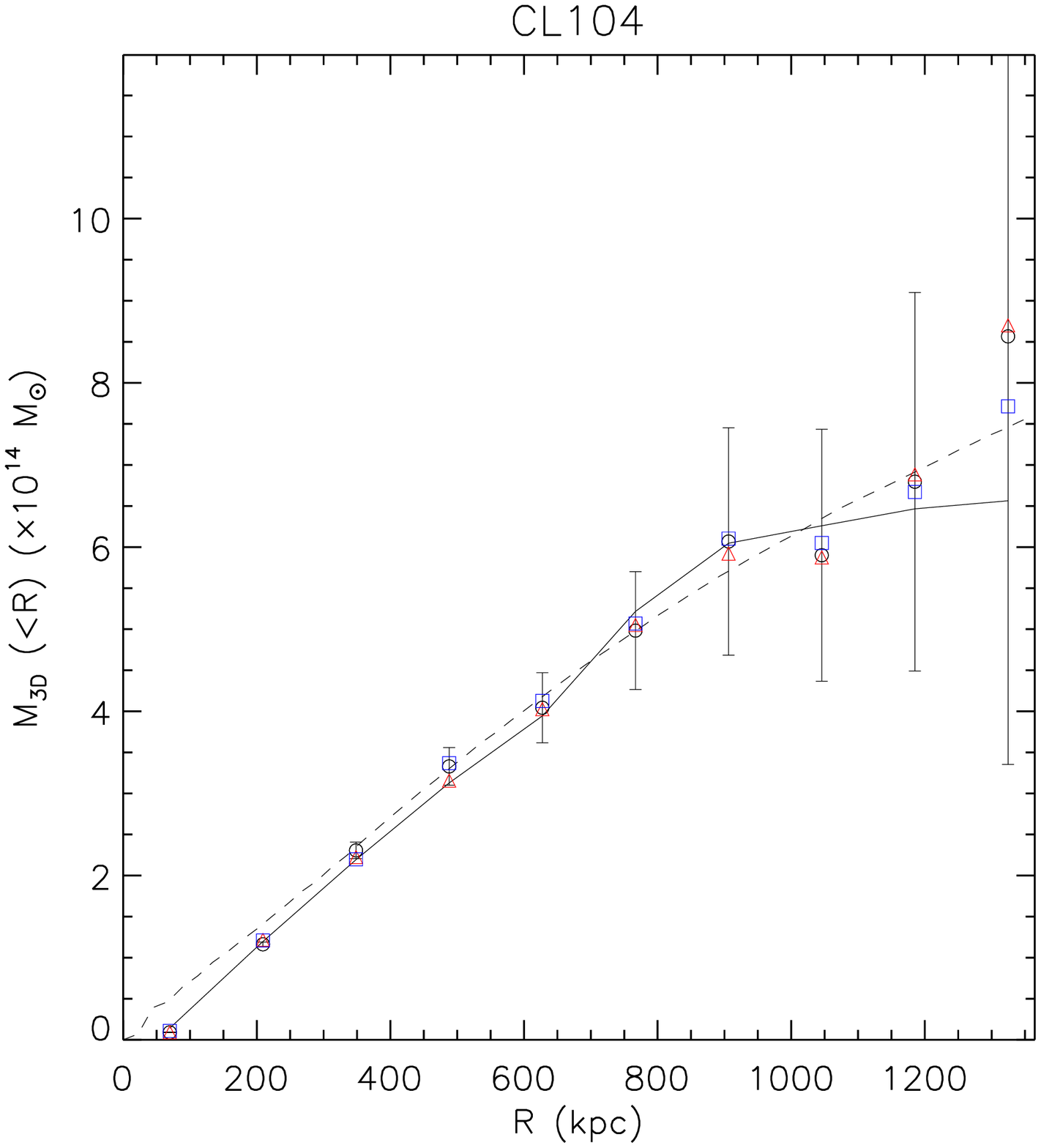,width=0.47\textwidth}
}
\caption[]{The recovery of 3D mass profile from mock CCAT analyses of CL101 and CL104. The solid line represents the hydrostatic mass using the true mass-weighted temperature and pressure, while the dashed one is the true spherically-averaged mass. The open points with errorbars are the recovered hydrostatic mass along the $z$ projection, while the triangles (squares) refer to the $x$ ($y$) projection (their errorbars are omitted for clarity).}
\label{ent246gtt3}
\end{center}
\end{figure*}

\section{Reconstruction of 3D Cluster Profiles}\label{phys11}
\subsection{Gas temperature, pressure and total mass}\label{phys11c}

We test the performance of our inverse method for reconstructing the desired 3D properties of the ICM using hydrodynamical simulations of galaxy clusters. Using the non-parametric method, we deproject ${\bf s}$ in order to infer ${\bf \epsilon}$ (\S \ref{apecdepte}), by rebinning the SZ brightness in circular annuli around the cluster center for each frequency and projection axis. Given the spectral signature of $\Psi(\nu;T({\bf r}),\beta({\bf r}))$, we then separate the physical variables $P({\bf r}),T({\bf r}), \beta({\bf r})$ by comparing the deprojected observed quantity ${\bf \epsilon}$ to its theoretical prediction. Errors have been evaluated via Monte-Carlo randomization, by generating correlated random numbers ${\bf p}$ with probability density function $f=f({\bf s},{\bf \sf C})$ where ${\bf s}$ is the expectation value and ${\bf \sf C}$ is the covariance. The final error estimate is accomplished by diagonalizing the covariance matrix ${\bf \sf C}$. In practice, by using its eigenvectors $V$, we rotate into a space in which the covariance matrix becomes ${\bf \sf C'}={\rm diag}({\sigma^2_1}',{\sigma^2_2}',...,{\sigma^2_p}')$, i.e. it is diagonal with eigenvalues ${\sigma^2_1}',{\sigma^2_2}',...,{\sigma^2_p}'$. We can thus treat the rotated parameters ${\bf p'}$ as independent and uncorrelated measurements with probability density $f=f({\bf s},{\bf \sf C'})$. To obtain the probability density function $f=f({\bf s},{\bf \sf C})$ we use the following relation:
\begin{equation}\label{cov2}
{\bf p}={\bf \sf V} \# {\bf p'}\ .
\end{equation}

Figure \ref{ent246g} shows the recovery of 3D temperature and pressure profiles from the mock CCAT analysis of CL101 and CL104 clusters viewed along three orthogonal projections. CL101 is a massive, dynamically active cluster, which experiences violent mergers at $z\sim 0.1$ and $z\sim 0.25$. CL104 is a similarly massive cluster, but with a more quiescent history. This cluster has not experienced a significant merger for the past 6~Gyrs, making it one of the most relaxed systems in the N07 sample. 

\begin{figure}
\begin{center}
\psfig{figure=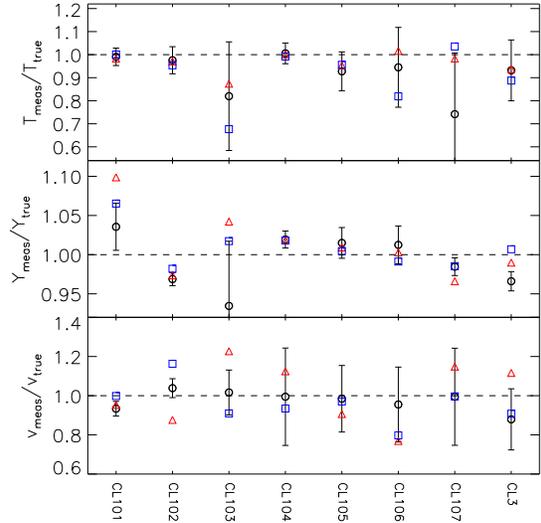,width=0.47\textwidth}
\caption[]{Upper panel: ratio between the measured and true mass-weighted temperature averaged inside a sphere of $R_{500}$ for the sample of simulated clusters. Middle panel: ratio between the measured and true Compton parameter integrated within $R_{500}$. Lower panel: ratio between the measured and true peculiar velocity along the line-of-sight, mass-weighted averaged inside a sphere of radius of $R_{500}$. For all the panels, the open points with errorbars refer to the $z$ projection, while the triangles (squares) refer to the $x$ ($y$) projection (errorbars are omitted for clarity).}
\label{entps3xkn3e}
\end{center}
\end{figure}

\begin{figure*}
\begin{center}
 \hbox{
\psfig{figure=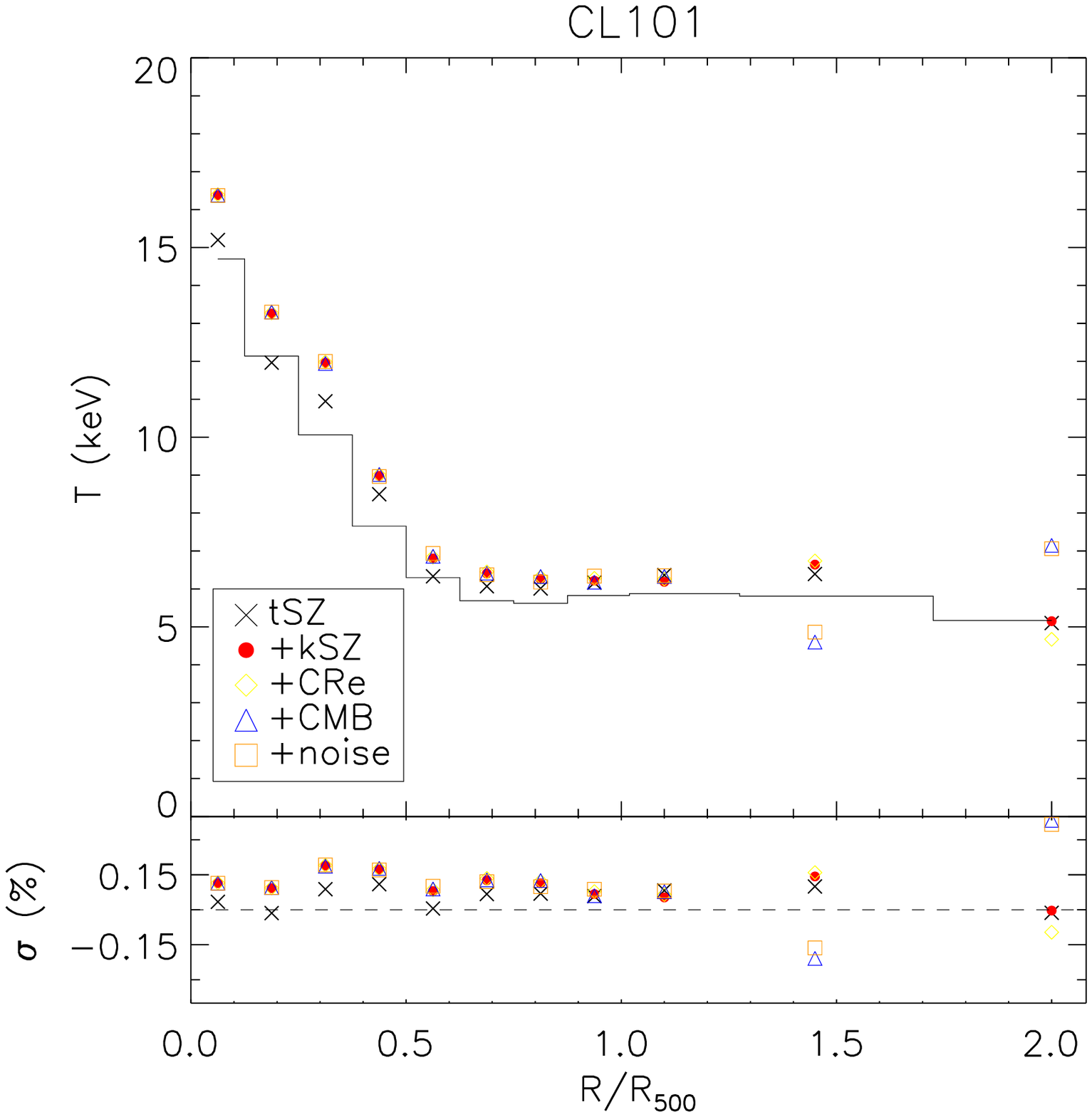,width=0.48\textwidth}
\psfig{figure=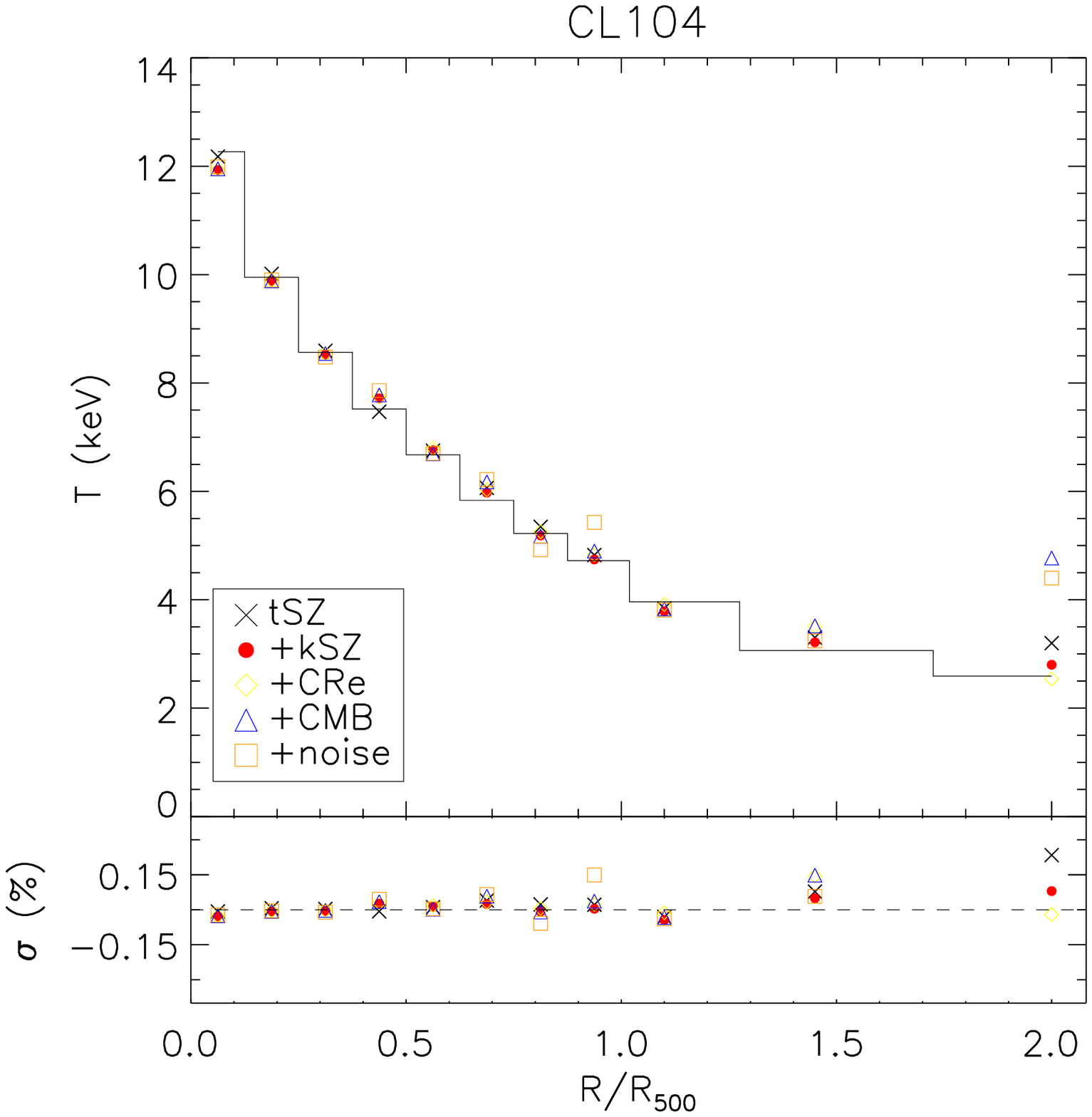,width=0.48\textwidth}
}
\caption[]{Reconstruction of 3D temperature profiles from the mock CCAT analyses of CL101 and CL104 viewed along the $z$ projection, in which we add contaminants to the tSZ signal one at a time, starting with the kSZ signal, the relativistic electrons (CRe), the primary CMB anisotropies, and the thermal noise in order. The different colored points reflect different astronomical and instrumental effects which degrade the recovery of profiles (see legend in the figures). The black-solid line represents the true mass-weighted temperature of the clusters in the $i$th spherical shell, whose width is graphically represented by the horizontal solid lines. We also report the fractional total scatter at the bottom of each panel.}
\label{entps3rkn3}
\end{center}
\end{figure*}

We find a remarkable agreement between the "measured" temperature (pressure) profiles and the true mass-weighted temperature (pressure) for the relaxed cluster CL104. We find a variation of order $5-15\%$ in both the recovered pressure and temperature profiles obtained for different projections of the same cluster. The recovery of the ICM profiles is still quite good for the unrelaxed cluster CL101 with the systematic bias at the level of $15-30$\%, due to the combination of the presence of substructures and the departure of the ICM distribution from the spherical symmetry. Moreover, internal bulk motions can introduce additional biases in the derived ICM profiles, especially in disturbed systems (e.g., characterized by the high values of $w$ in Table~\ref{tabdon}). Note that via our non-parametric method the full information about the ICM profile is preserved in converting the SZ brightness profile to 3D temperature and pressure profiles. 

We also assess the impact of the primary CMB anisotropies on the derived ICM profiles, by comparing the recovered profiles from the maps with and without CMB anisotropies, while always including the thermal noise. We find that the bias due to the primary CMB anisotropies is small in the cluster cores, but they become comparable to that of the thermal noise in the outer radii ($r\gesssim R_{500}$).
        
In Figure \ref{ent246gtt3}, we show the mass profiles obtained by applying the hydrostatic equilibrium equation on the temperature and pressure profiles derived from the mock CCAT analysis. Similarly, we also compute the hydrostatic mass profiles using the mass-weighted temperature and pressure measured directly in simulations. We then compare these hydrostatic mass profiles to the true mass measured directly from simulations in order to assess the bias in the recovered mass profiles. We find that the multifrequency CCAT SZ analysis of individual cluster can recover the hydrostatic mass profiles very accurately in the entire radial range for the relaxed cluster. Note, however, that the hydrostatic mass is biased low compared to the true mass by $5-10\%$ (N07b), primarily due to the violation of the hydrostatic condition even in the inner regions of a very relaxed cluster (CL104). In the unrelaxed cluster (CL101), the bias is larger especially in the outer regions where the departure from the hydrostatic equilibrium is even more significant. Note that the bias in the hydrostatic mass could be corrected if we could measure the level of bulk and turbulent gas motions in clusters (e.g., by kSZ effect) and account for the non-thermal pressure provided by these components \citep{lau2009,nelson2012}.

In Figure \ref{entps3xkn3e} we demonstrate the recovery of the global properties of the ICM, by showing the ratios of the measured and true mass-weighted temperature, integrated Compton parameter ($Y_{\rm SZ}$), mass-weighted line-of-sight peculiar velocity within a sphere of radius $R_{500}$ for the sample of simulated clusters.  We find a non-negligible scatter in both the reconstructed temperature and $Y_{\rm SZ}$. For 16 hours of CCAT observations, the total scatters (intrinsic plus statistical measurement errors) are $\sim$13\% for the mass-weighted temperature and 4\% for $Y_{\rm SZ}$, while the intrinsic scatters (variance of the data points without the measurement errors) are relatively small ($\sim4$\% and $\sim3$\%, respectively), primarily due to substructures and triaxiality of the ICM distribution.  

In Figure \ref{entps3rkn3}, we assess the impact of various sources of uncertainties by adding contaminants to the tSZ signal one at a time, starting with the kSZ signal, the relativistic electrons, the primary CMB anisotropies, and the thermal noise in order. We find that the temperature profile is recovered very well for CL104. The recovery is not as accurate for CL101, in which the recovered temperature could be biased by $5-25$\% at any given radius. In absence of other sources of noise (i.e., using the tSZ data alone), the reconstructed 3D temperature profiles are biased by only $\sim4\%$ for CL104 and $\sim9\%$ for CL101. Biases are generally larger for the unrelaxed systems that exhibit more significant departures from the spherical symmetry indicated in Table~\ref{tabdon}. At large radii ($r \gesssim R_{500}$), we find that the primary CMB anisotropies are one of the main sources of systematic uncertainties in the 3D temperature reconstruction. For example, at $r=3\,R_{500}$ the bias due to the primary CMB anisotropies is $\sim$ 10\% of the SZ signal at 93~GHz. In the inner regions ($r \lesssim R_{500}$), the primary CMB anisotropies introduce the scatter at the level of 7\% and 10\% for CL104 and CL101, respectively. Our analyses indicate that the relativistic electrons, peculiar velocity and thermal noise have little effect on the recovered 3D temperature of a hot ($T_{\rm mw} \gesssim 3$~keV) galaxy clusters.

Recent SZ surveys showed that clusters at higher redshift (e.g. $z\gesssim 0.6$) seem more disturbed than their low-redshift counterparts \citep{vanderlinde2010,planck2011}. To assess possible differences expected for high-redshift clusters, we also analyzed the outputs of the simulated galaxy clusters at $z=0.6$. First, higher-redshift clusters are projected over a smaller area of the sky: this translates into a poorer spatial resolution in the measured three-dimensional parameters and larger measurement errors (up to 20-30\%) due to the larger impact of the CMB anisotropies.  Second, the high-redshift clusters are more disturbed on average: this leads to larger systematic errors as well.

\subsection{Cluster peculiar velocities}\label{phys11d}
Galaxy clusters are excellent tracers of the large-scale velocity field in the universe. The kSZ effect, due to the relative motion of the cluster with respect to the rest frame of the CMB, enables direct measurements of the cluster peculiar velocity independently of the cluster distance. In principle, this technique can be used to measure deviations of the cluster's motion from the Hubble flow with high precision at any redshift. This is in contrast to the standard candle (e.g., SNe) whose measurement errors increase monotonically with redshift. In this section we evaluate how well we can measure the peculiar velocity of clusters using multifrequency SZ observations. 

\begin{figure}
\begin{center}
\psfig{figure=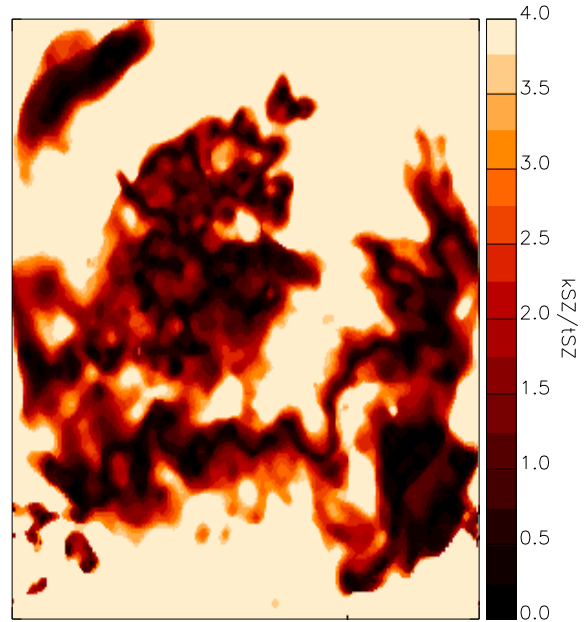,width=0.5\textwidth}
\caption[]{Absolute value of the ratio between the kinetic and thermal SZ effect via an observation of CL101 with a receiver of bandwidth equal to 7.5~GHz and centered on 218~GHz. We take a Hanning window for the response of the radio receiver and we projected the datacube along the $z$ direction. The size of the region shown in the figure is about 5.7~Mpc with our cosmological parameters, with a pixel size of about 22~kpc, and the region is centered on the minimum of cluster potential.}
\label{ent2t8}
\end{center}
\end{figure}

In practice, there are a number of complications for the kSZ measurements of the cluster peculiar velocity: (i) the kSZ effect is much smaller than the tSZ effect and is entangled with other contributions to Comptonization of the CMB, such as energetic non-thermal electrons: difficulties might arise in separating these different spectral signatures in SZ signal $P({\bf r})\,\Psi(\nu;T({\bf r}),\beta({\bf r}))$ via multifrequency SZ data; (ii) the blackbody kSZ distortion is degenerate with the spectrum the primordial CMB anisotropies, which is also a blackbody; (iii) the observed kinematic SZ signal gets significant contributions from the internal gas motions in galaxy clusters, limiting the accuracy of the peculiar velocity measurement at the level of $50-100$~km/s for individual clusters \citep{nagai2003}. 

Concerning the first point, the cluster peculiar velocity is often estimated from the kSZ maps alone, by focusing on observations at the crossover frequency at 218~GHz where the tSZ emission vanishes. However, the exact value of the crossover frequency depends weakly on the electron spectrum \citep[gas temperature and density,][]{birkinshaw1999}. Moreover, real observations are obtained with receivers with a finite bandwidth typically of order a few GHz. The kSZ observations are contaminated by other signals, including tSZ emission and non-thermal SZ from relativistic electrons. This is illustrated in Figure \ref{ent2t8}, which shows the ratio between the kinetic and thermal SZ effect obtained by receivers with bandwidth of 7.5~GHz and centered on 218~GHz. We can see that the thermal and kinetic SZ signals are comparable in intensity; i.e. the tSZ effect does not fully vanish because of the finite width of the bandwidth and the dependence of the crossover frequency on the gas temperature and density. 

\begin{figure*}
\begin{center}
\hbox{
\psfig{figure=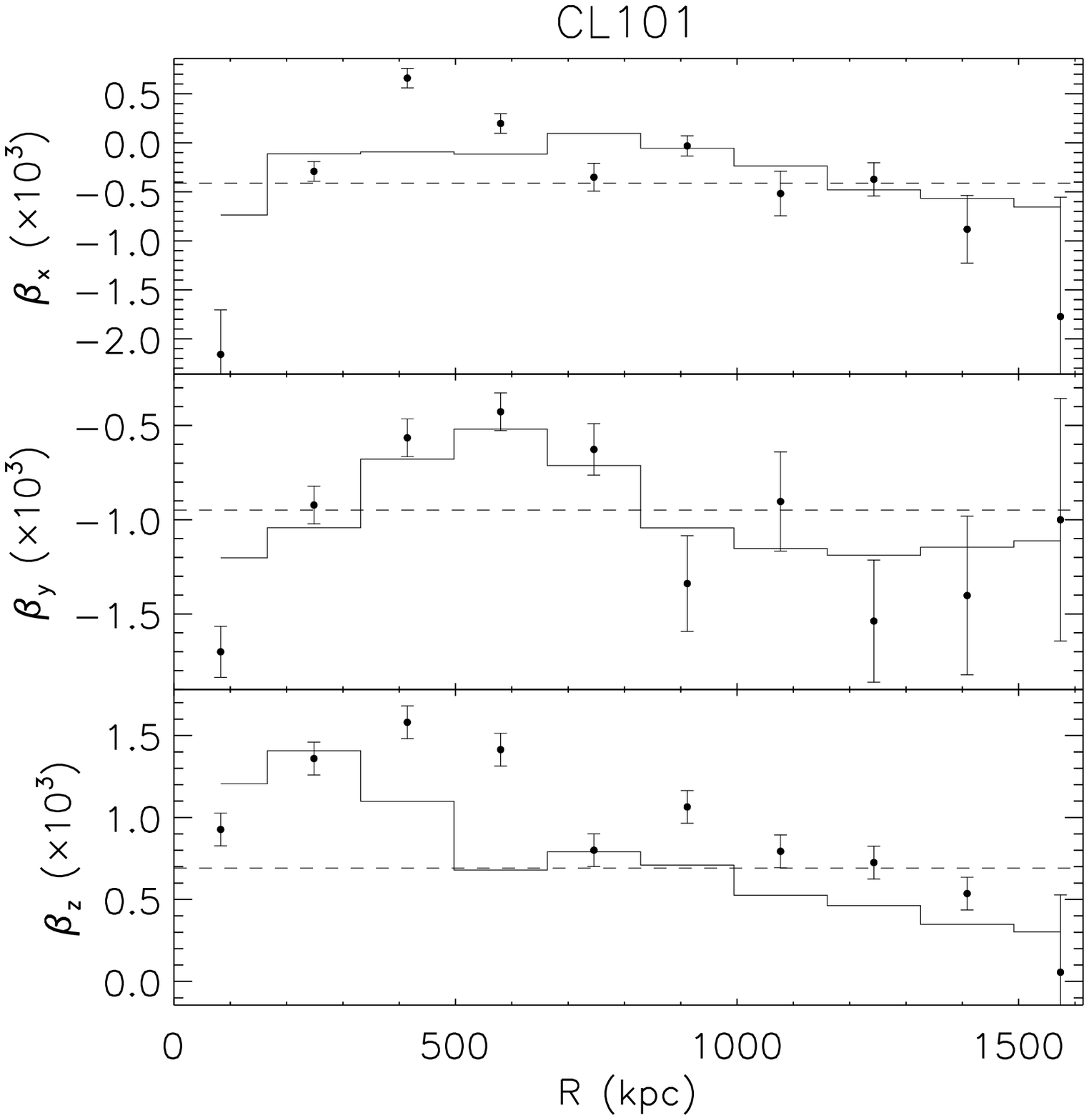,width=0.48\textwidth}
\psfig{figure=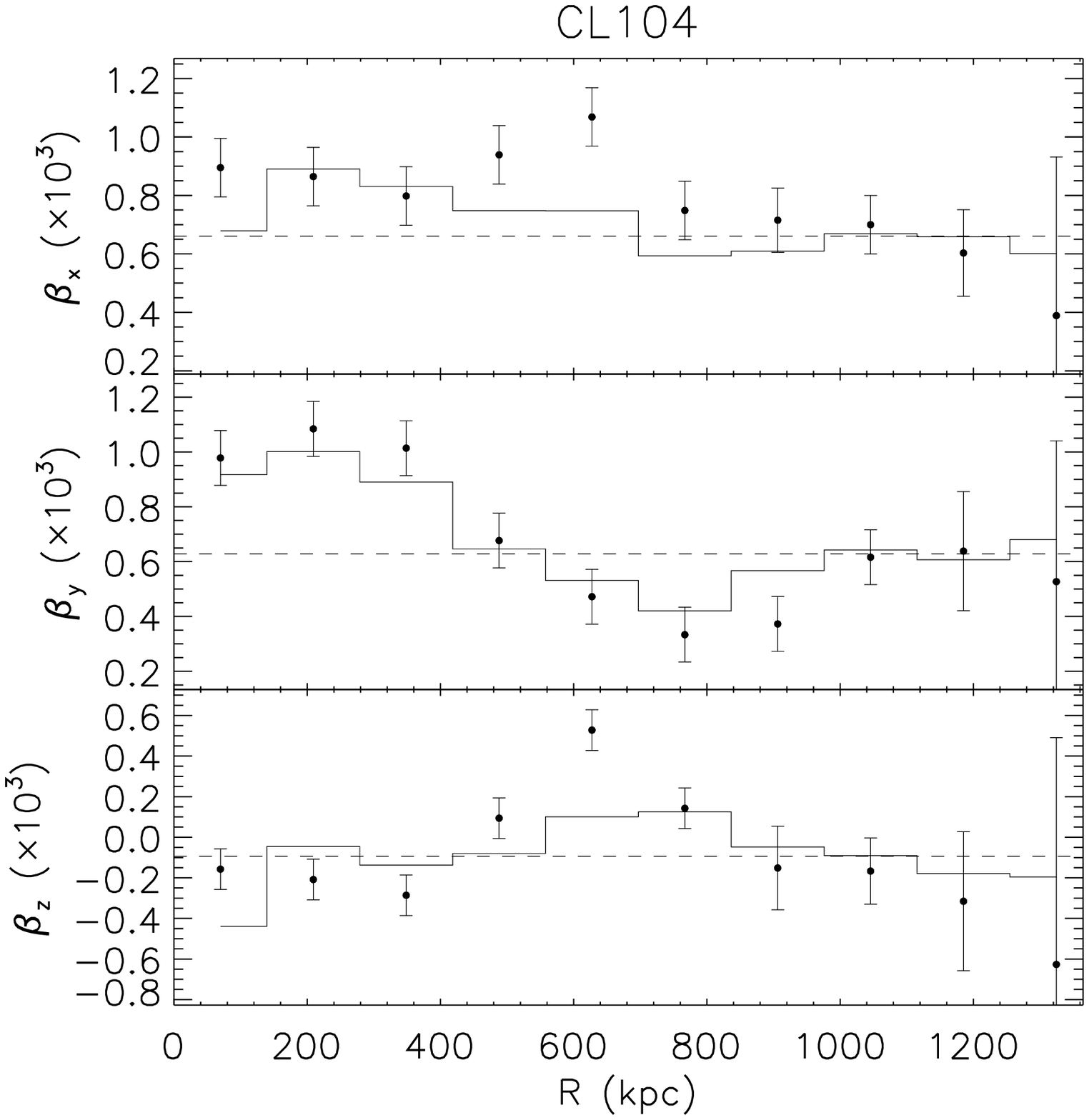,width=0.48\textwidth}
}
\caption[]{The recovered spherically-averaged streaming velocity profiles $\beta({\bf r})$ from mock CCAT analyses of simulated clusters viewed along three orthogonal projections ($x$, $y$, $z$). The black-solid lines represent the true spherically averaged mass-weighted streaming velocity of the clusters in the $i$th spherical shell, whose width is graphically represented by the horizontal solid lines. The points with errorbars are the streaming velocities retrieved from the mock analysis. The dashed lines refer to the true mass-weighted streaming velocity averaged inside a circular aperture of radius equal to the boundary of the observation.}
\label{ent246gf}
\end{center}
\end{figure*}

To address these issues, we analyze mock CCAT simulations in order to assess how well we can measure the peculiar velocity of galaxy clusters. Figure \ref{ent246gf} shows the reconstructed streaming velocity profile $\beta({\bf r})$ as a function of the cluster-centric radius. We find that the velocity radial profile $\beta({\bf r})$ in the ICM matches the true spherically-averaged cluster peculiar velocity. Our retrieved streaming velocity profiles suggest that systematic uncertainties in the measured $\beta({\bf r})$ is typically of order of 100 km/s, but there are occasional outliers with bias of order a factor 2 in a certain projection axis of disturbed clusters. For example, the largest bias is found in CL106, which is a disturbed system with a large value of $w$ (Table~\ref{tabdon}). Therefore, one way to control such a large bias is to consider only relaxed systems with a low value of $w$ and average the peculiar velocity out to the SZ boundary. Note also that the biases are present even in relaxed systems due to frequent minor mergers that continuously perturb the clusters gas atmosphere \citep[][]{vazza2009,lau2009}.

Next, we consider the mass-weighted peculiar velocity averaged within a circular aperture. Averaging out within a fixed radius has two advantages: (i) the mass-weighted peculiar velocity is more easily obtainable with SZ observations with moderate angular resolution; (ii) the bias in the measured mass-weighted streaming velocity tends to be smaller than the differential radial profiles $\beta({\bf r})$. In Figure \ref{entps3xkn3e} we present the ratios between the measured and true peculiar velocities obtained along three orthogonal projections, where the recovered velocity is mass-weighted averaged inside a circular aperture of radius equal to the boundary of the observation. For most of the objects, the average peculiar velocities derived from different projection axes are consistent within statistical errors of order $10-100$~km/s (corresponding to a fractional uncertainty of $\sim10-15$\%). This dispersion sets a theoretical lower limit on the accuracy with which cluster peculiar velocity can be measured for an individual cluster. The bias in the averaged peculiar velocity mainly arises from: (i) internal motions of the ICM, which gives an systematic error of around $50-100$~km/s \citep{nagai2003}; (ii) triaxiality, which introduces an systematic error of $\lesssim50$~km/s; (iii) systematic intensity change $\Delta I_{\rm rel}(\nu)$ due to relativistic electrons, which gives an systematic error of $\lesssim20$~km/s. These errors are also largely independent of the cluster redshift, in contrast to galaxy peculiar velocity surveys. Therefore, kSZ measurements of the cluster peculiar velocities might serve as promising tools for studying cosmology, including the expansion history of the universe, the matter density of the universe, the primordial power spectrum normalization, and the dark energy equation of state \citep{bhattacharya2007}.

\section{Discussions and Conclusions}\label{conclusion33}

In this paper we presented a non-parametric inversion method for reconstructing 3D temperature and pressure profiles of the ICM from multifrequency SZ observations. In order to assess the accuracy to which the 3D physical parameters can be recovered from observations, we analyzed mock Cerro Chajnantor Atacama Telescope (CCAT) observations of simulated clusters, while taking into account several sources of uncertainties associated with instrumental effects and astrophysical foregrounds.

We showed that our method enables the recovery of the 3D ICM temperature and pressure profiles out to large radius ($r\approx R_{500}$) of a hot ($T_X\gesssim 3$~keV) cluster, while retaining full information about the gas distribution. Accurate recovery of the ICM pressure and temperature profiles, in turn, enables accurate determination of the mass-weighted temperature, integrated SZ signal ($Y_{\rm SZ}$), and the cluster mass profiles assuming the hydrostatic equilibrium of the gas in the cluster potential.  We found that the bias in the reconstructed 3D ICM and mass profiles are dominated by substructures and asphericity of the cluster gas distribution, while the contaminations by thermal noise, relativistic electrons and the CMB anisotropies are small.  Note also that the hydrostatic mass is a biased estimate of the true cluster mass, due to the presence of non-thermal pressure provided by bulk and turbulent gas motions.  A joint analysis of SZ and lensing measurements may help constrain the biases due to triaxial shape and non-thermal pressure further \citep{morandi2010a,morandi2011b,morandi2012a}.

We also assessed the accuracy to which the cluster's peculiar velocity could be measured with upcoming multifrequency observations of the kSZ effect. We find that there is a non-negligible total scatter of $\sim 10-15$\% for individual clusters. For 16h hours of CCAT observation, the measurement errors are still dominated by statistical uncertainties, instead of systematic uncertainties. The systematic bias in the averaged peculiar velocity mainly arises from: (i) internal motions of the ICM ($\approx 50-100$~km/s); (ii) triaxiality ($\lesssim50$~km/s); (iii) systematic intensity change $\Delta I_{\rm rel}(\nu)$ due to relativistic electrons ($\lesssim20$~km/s). Thus, CCAT measurements of the cluster peculiar velocities via kSZ effect should serve as unique and potentially powerful probes of the dynamics of the universe and cosmological parameters.

The inversion method presented here should be a valuable tool for analyzing multifrequency SZ observations of galaxy clusters. While the SZ effect has been used mainly for constraining the electronic pressure to date, high-resolution, multifrequency SZ observations (e.g., by CCAT) will enable detailed characterization of the temperature and velocity structures in the ICM, without follow-up X-ray observations. SZ observations are especially unique and powerful in probing the outskirts of high-redshift clusters and complementary to X-ray observations which are sensitive to the inner regions of nearby clusters. In this sense, CCAT SZ observations are complementary to the upcoming eROSITA X-ray cluster survey\footnote{{http://www.mpe.mpg.de/eROSITA}} with comparable angular resolutions. Furthermore, CCAT SZ observations will provide unique constraints on the cluster peculiar velocities and possibly internal gas flows especially in the outskirts of high-redshift clusters, which are difficult to measure using the upcoming Astro-H mission\footnote{{http://astro-h.isas.jaxa.jp -- a Japanese-US space based X-ray telescope scheduled to launch in 2014.}} via shifting and broadening of X-ray spectral lines. The present work is a step forward toward robust reconstruction of the 3D ICM, mass, and streaming velocity profiles for a sample of clusters and promises to help advance cosmological studies based on the cluster growth over cosmic time in the coming decade.

\section*{acknowledgements}
We acknowledge Kaustuv Basu, Jens Chluba, Irina Dvorkin, Shimon Meir, Yoel Rephaeli, Tony Mroczkowski and Jack Sayers for useful discussions. This work was supported in part by the US Department of Energy through Grant DE-FG02-91ER40681 and Purdue University. DN acknowledges support from NSF grant AST-1009811, NASA ATP grant NNX11AE07G, NASA Chandra Theory grant GO213004B, Research Corporation, and by Yale University. This work was supported in part by the facilities and staff of the Yale University Faculty of Arts and Sciences High Performance Computing Center. 


\newcommand{\noopsort}[1]{}

\end{document}